\documentclass{article}

\usepackage{arxiv}
\usepackage{xcolor}
\usepackage[textsize=footnotesize]{todonotes}
\usepackage[font=small,skip=0pt]{caption}
\usepackage{enumitem}
\usepackage{float}
\usepackage{textpos}
\usepackage[style=vancouver]{biblatex}
\usepackage{tabularx,longtable,multirow,subfigure,caption}%hangcaption
\usepackage{tabularray}
\usepackage{fncylab} %formatting of labels
\usepackage{fancyhdr} % page layout
\usepackage{url} % URLs
\usepackage[english]{babel}
\usepackage{amsmath}
\usepackage{graphicx}
\usepackage{grffile}
\usepackage{dsfont}
\usepackage{epstopdf} % automatically replace .eps with .pdf in graphics
\usepackage{array}
\usepackage{latexsym}
\usepackage{mathtools}
\usepackage{amsmath}
\usepackage{listings}

\usepackage{xcolor}

\definecolor{codegreen}{rgb}{0,0.6,0}
\definecolor{codegray}{rgb}{0.5,0.5,0.5}
\definecolor{codepurple}{rgb}{0.58,0,0.82}
\definecolor{backcolour}{rgb}{0.95,0.95,0.92}

\lstdefinestyle{mystyle}{
    backgroundcolor=\color{backcolour},   
    commentstyle=\color{codegreen},
    keywordstyle=\color{magenta},
    numberstyle=\tiny\color{codegray},
    stringstyle=\color{codepurple},
    basicstyle=\ttfamily\footnotesize,
    breakatwhitespace=false,         
    breaklines=true,                 
    captionpos=b,                    
    keepspaces=true,                 
    numbers=left,                    
    numbersep=5pt,                  
    showspaces=false,                
    showstringspaces=false,
    showtabs=false,                  
    tabsize=2
}

\title{Benchmarking Specialized Databases for High-frequency Data}

\author{ Fazl Barez \\
        Edinburgh Centre for Robotics \\
        The University of Edinburgh\\
        \texttt{f.barez@ed.ac.uk}\\
    \And
    Paul Bilokon \\
	Department of Computing \\
	Imperial College London \\
	\texttt{paul.bilokon@imperial.ac.uk} \\
	\And
	Ruijie Xiong \\
	Department of Computing \\
	Imperial College London \\
	\texttt{ruijie.xiong21@imperial.ac.uk} \\}

\begin{document}

\maketitle

%%%%%%%%%%%%%%%%%%%%%%%%%%%%%%%%%%%%
\begin{abstract}
This paper presents a benchmarking suite designed for the evaluation and comparison of time series databases for high-frequency data, with a focus on financial applications. The proposed suite comprises of four specialized databases: ClickHouse, InfluxDB, kdb+ and TimescaleDB. The results from the suite demonstrate that kdb+ has the highest performance amongst the tested databases, while also highlighting the strengths and weaknesses of each of the databases. The benchmarking suite was designed to provide an objective measure of the performance of these databases as well as to compare  their capabilities for different types of data. This provides valuable insights into the suitability of different time series databases for different use cases and provides benchmarks that can be used to inform system design decisions.
\end{abstract}

%%%%%%%%%%%%%%%%%%%%%%%%%%%%%%%%%%%%
\section{Introduction}
\subsection{Motivation}

In the era of big data, various industries such as finance and technology generate large volumes of time series data every day. \emph{Time series} data are characterized by being large in size and updated more frequently than classical relational data, meaning that new data points are continuously generated while old data are rarely updated. As a result, traditional database systems require optimization to enable efficient storage and operation of these time series data. Specialized time series databases are needed to meet the stringent requirements of read/write throughput and latency in real-world scenarios. 

Common specialized databases for high-frequency data include InfluxDB and Kdb+. Both databases offer a range of features for time series data storage, management and analysis. InfluxDB is used for storing and processing time series data and provides a powerful query language for data retrieval. Kdb+ is a columnar database and offers support for high-frequency data analytics through its in-memory databases, query language, and the analytics platform.

specialized databases are proving to be essential in catering to the specific features of time series data, and are becoming increasingly more popular for applications requiring high-frequency data management and analysis.

Choosing the appropriate database for a company requires benchmarks to effectively evaluate and compare different time series databases. Traditional relational database benchmarks, such as the TPC benchmarks~\parencite{9458659}, are not suitable for large-scale time series data applications. Thus, the STAC-M3 benchmarking suite~\parencite{STAC-M3} was developed for financial tick data analysis, but has been applied to a limited number of databases, such as kdb+ and eXtremeDB.

Cooper et al.~\parencite{cooper2010benchmarking} proposed the YCSB open source benchmarking suite for big data applications, though not applicable to typical workloads for time series databases. To address this gap, we present a specialized open source benchmarking suite applicable to time series data used in finance. The workloads are simulated using cryptocurrency exchange data, and the suite is applied to ClickHouse, InfluxDB, kdb+, and TimescaleDB for evaluation.

%%%%%%%%%%%%%%%%%%%%%%%%%%%%%%%%%%%%
\section{Literature Review}
\subsection{Time Series Databases}

\emph{Time series data} are characterized by immutability and continuous, dynamic updates, and are often large in size. These data, which often result from financial applications, differ from traditional data (e.g., relational tables), and are utilized for analysis and data mining purposes and are often large~\parencite{fu2011review}. Immutability is also an attribute of time series data: new data points are always appended and recorded data is rarely modified.

Traditional databases have redundant features and are not optimized for time series data operations, such as arbitrary update or deletion of stored data~\parencite{shafer2013specialized}. Moreover, time series data analysis typically involves data in a particular temporal range rather than the whole dataset, and recent data is typically accessed more frequently than older data~\parencite{bitincka2010optimizing}.

Recent research has suggested that traditional databases can be challenging to scale to time series data due to its often large size. To improve storage efficiency, compression and deletion strategies can be implemented, and modifications can be applied to compression algorithms to optimize data with distinct characteristics~\parencite{przymus2014dynamic}. Additionally, preprocessing and summarization techniques could be utilized to expedite query processing time~\parencite{shafer2013specialized}. 

LittleTable~\parencite{rhea2017littletable} is a distributed relational database developed by Cisco Meraki to store time series data for device monitoring and analysis. It is optimized to partition data according to timestamps and attributes, store data of the same hierarchy collected during the same time range together on disk, and implement a configurable time-to-live (TTL) policy to delete outdated data automatically. This reduces the need to store the large amount of data in main memory. Additionally, its single-writer and  append-only strategy eliminates the need for locking and updating of previous data. LittleTable is also fault-tolerant, as the data is served from a server geographically closest to the user and can be replaced by a spare server that can reconstruct the database from write-ahead logs in case of main server failure. 

Gorilla \parencite{pelkonen2015gorilla}, Facebook's highly-available, in-memory time series database, uses a compression scheme based on delta-of-deltas to reduce data storage by an average of 12x. This approach transforms the collected data points by finding the difference between the current data point and the previous two points, resulting in a small value even when the difference in data points is insignificant. Gorilla does not prioritize consistency like other relational databases, in order to focus on write availability and  read performance; as time series data is usually used for aggregated analysis, any lost data due to inconsistency has minimal impact on the overall result.

Monarch \parencite{adams2020monarch}, developed by Google, is a multi-tenant in-memory time series database that supports the monitoring and analysis of a variety of services. The schematized and relational model of Monarch contains various data types and a rich query language to address the varying needs of its users. Data is stored in partitions based on the lexicographical order of the target columns, which are composed of the target columns and metric columns associated with each time series, in order to enable efficient access.  In order to reduce the amount of in-memory storage required, strategies such as shared timestamp sequences and delta encoding of complex types are employed. Additionally, Monarch provides collection aggregations, which will aggregate the data according to user-defined configurations before it is stored in-memory. Monarch is designed for high write availability, justifying the dropping of delayed writes in the face of network congestion in favour of more recent requests. High read availability is ensured by allowing queries to return partial data and alerting users when some servers become unresponsive via the soft query deadline.

\subsubsection{kdb+}\label{sec:kdb}

Kdb+ \parencite{taylor_2022} is a commercial in-memory time series database developed by KX Systems. To support fast ingestion and immediate query of data, kdb+ migrates data in the hierarchy of Realtime Database (RDB), Intraday Database (IDB), and Historical Database (HDB) \parencite{kdb_arch}. Data is first stored in the in-memory RDB, and data accumulated during a user-configured time interval is moved to the temporary on-disk IDB. Data in IDB is further organized and sorted and later stored in the permanent on-disk HDB. Kdb+ stores the data in columnar representation as data in the same column are stored together, opposing to the row-oriented architecture of traditional relational databases. This greatly reduces the amount of data to be accessed to serve queries because queries are often targeted to certain attributes. Kdb+ relies on a built-in programming and query language that has a relatively small size of around 800 kb, which could be stored in the L1/2 cache of the CPU for high-speed access \parencite{kx_blog}. Additionally, kdb+ performs computation and filtering in the database, reducing the overhead of sending data over the network for computation.

Kdb+ has shown excellent performance and has a wide array of applications in the industry. Kdb+ has been tested to be 29x faster than Cassandra and 38x faster than MongoDB \parencite{kdb_comparision}. Kdb+ holds the record for 10 out of 17 benchmarks in the Antuco suite and 9 out of 10 benchmarks for the Kanaga suite of the STAC-M3 benchmark, which is the financial services industry standard for time series data analytics benchmarking \parencite{kdb_benchmark}. Cobalt has built its post-trade infrastructure system using kdb+, which supports throughput of 1 million trades per second and reduces the cost by up to 80\% \parencite{kdb_cobalt}. A global investment bank utilizes Google Cloud and kdb+ to build a scalable database for real-time and historical market data analysis, processing 1 TB of data every day \parencite{kdb_ib}. A fintech startup that incorporated AWS S3 with kdb+ achieved 10x performance with 20\% of the cost compared to using SSD \parencite{kdb_fintech}.

\subsubsubsection{In-Memory Database}

Due to the reduction of main memory storage device costs in recent years, it has become possible to use in-memory databases for enterprise applications. In-memory databases are databases where the data mainly resides in the main memory rather than the disk \parencite{plattner2012memory}. This allows for functionalities such as real-time availability and analysis of data, as main memory access is orders of magnitude faster than disk access. They also require non-volatile storage for logging and recovery purposes to ensure fault tolerance \parencite{plattner2013course}. It is also necessary for in-memory databases to implement efficient compression techniques to reduce storage without slowing down the performance. In-memory databases can also exploit multi-core CPUs to achieve parallelisation by partitioning the dataset according to processors.

\begin{figure}[H]
\centering
\includegraphics[width=\textwidth]{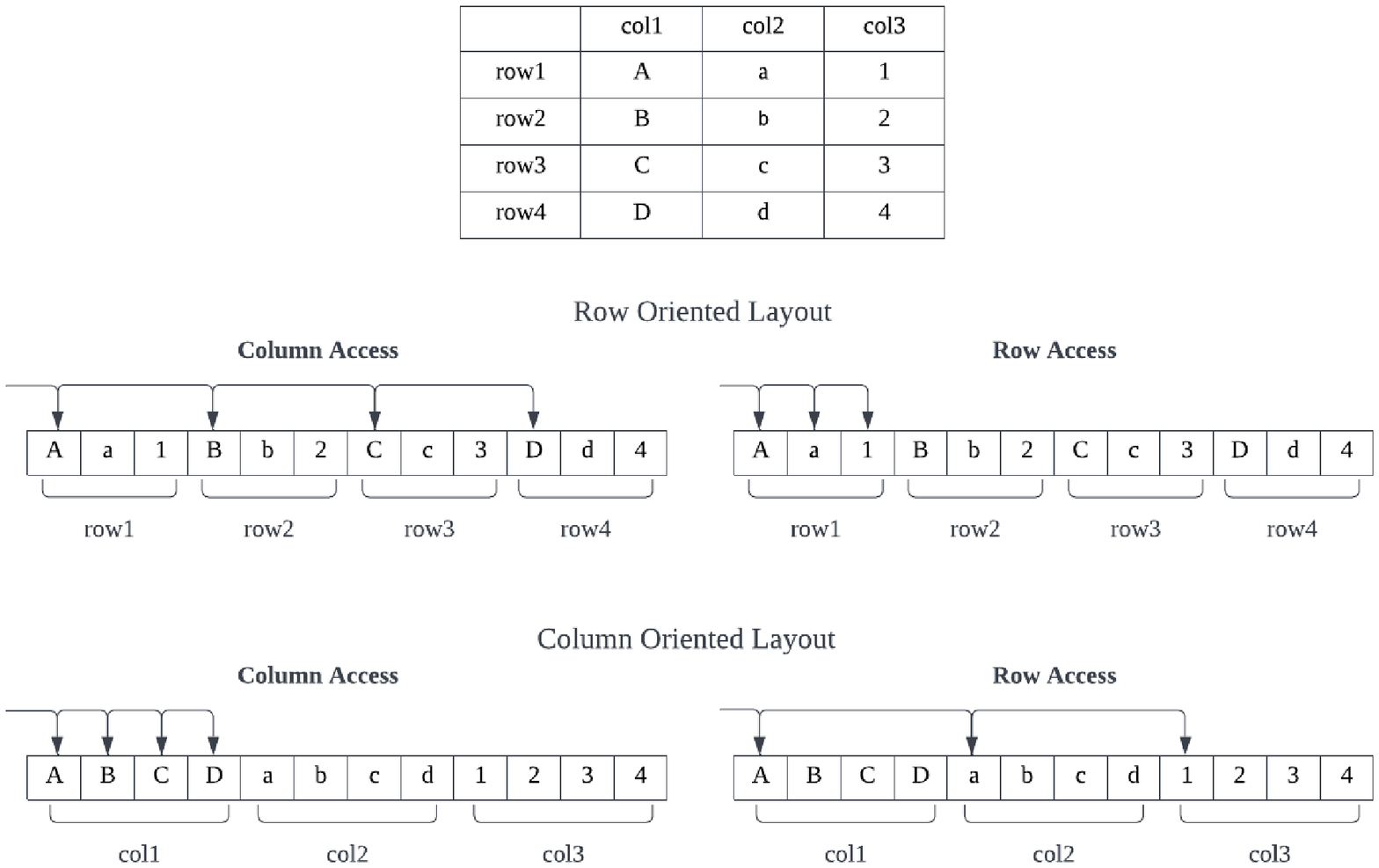} 
\caption{Column Oriented Layout}
\label{column layout}
\end{figure}

\subsubsubsection{Column Oriented Layout}

kdb+ uses a column-oriented data layout rather than the traditional row-oriented data layout of traditional relational databases \parencite{qformortals}. Figure \ref{column layout} compares the storage and access of these two storage models in memory. In kdb+ data of the same column is stored together and columns may be distributed across the storage, while in SQL data of the same row is stored together. As shown in Figure \ref{column layout}, the row access in row-oriented layout and column access in column-oriented layout performs sequential access of memory, which would best benefit from memory locality \parencite{tinnefeld2015building}. The row-oriented layout is suitable for Online Transaction Processing (OLTP) workloads which focus on transaction processing and operate on rows. Online Analytical Processing (OLAP) workloads that are mostly read operations and query over table attributes are more beneficial from a column-based layout. These OLAP access patterns are common in the use of time series databases.

\subsubsubsection{q Programming Language}

q is the programming language designed for kdb+, which is a strongly functional language optimized for expressive and effective database management, as opposed to traditional generalized languages such as C++, Java, and Python \parencite{qformortals}. q programs are interpreted from the scripts at run time rather than compiled. Benefiting from the concise syntax and small size, data and functions can reside in memory during execution, which leads to fast deployment and testing. Tables in q are first-class entities that are stored in memory like other data structures such as lists and dictionaries. Tables are implemented as ordered column lists, where each list occupies contiguous memory, optimizing for sequential access to exploit memory locality.

q offers many specialized functionalities for database operations and management \parencite{novotny2019machine}. q-SQL is a rich table querying syntax of q which is similar to SQL. Tables can be splayed, which means that they are divided vertically along the columns and stored as separate files on disk. Splayed tables can be further partitioned to be divided horizontally. Python or R could be combined with q as preprocessing or visualization tools. q has a wide array of applications in statistical analysis and machine learning with big data due to its efficient operations and fast speed.

\subsubsubsection{kdb+/tick} \label{sec:kdb tick}
\begin{figure}[H]
\centering
\includegraphics[width=\textwidth]{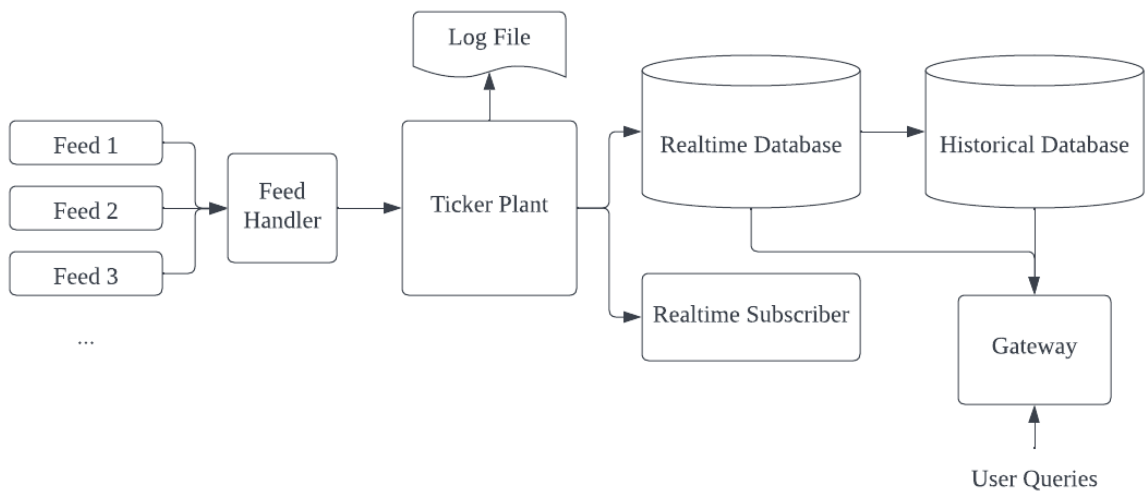} 
\caption{kdb+/tick Architecture}
\label{kdb tick}
\end{figure}

Figure \ref{kdb tick} demonstrates the kdb+/tick architecture for the capture, process, and analysis of real-time and historical data \parencite{kdb_arch}. The data feed come from trading venues and market data vendors such as Bloomberg. The feed handler would parse and extract the data from feed-specific formats to kdb+ standard format. When the ticker plant first receives the message, it writes the data to the log file for failure recovery. After that, the data is published to the real-time subscriber and saved to the real-time database in memory. The real-time subscriber could perform real-time processing and analysis upon receiving the data. At the end of the day, the intraday data in the real-time database in memory is persisted to the historical database on disk, the real-time database is also flushed along with the log file to accept new data the next day. The gateway is used to handle user queries to the database and return the query results. The result may require the data from the real-time database or the historical database, or it could be a combination of both.

\subsubsection{InfluxDB}

InfluxDB \parencite{influxdb} is an open-source database optimized for time series data developed by InfluxData. The data model and schema design of InfluxDB are different from traditional databases for storing time series data. The dataset is horizontally partitioned into shards that contain the data in a specific time interval according to the retention policy. InfluxDB balances the performance for write and read operations by caching and compaction techniques. Some tradeoffs are made to ensure scalability and availability. The Update and Delete functionalities in traditional CRUD are greatly restricted and cross table joins are not supported in InfluxDB's schema-less design \parencite{influxdb_tradeoffs}. Queries can return outdated or incomplete results to guarantee availability.

InfluxDB has been shown to have 4.5x write throughput, 2.1x less disk space, and up to 45x faster query res,ponse time than Creal-time \parencite{influxdb_Cassandra}, and the benchmarks compared to MongoDB are 2.4x, 20x and 5.7x respectively \parencite{influxdb_MongoDB}. CapitalOne built a highly-available and fault tolerant system for storing and analyzing their business, infrastructure, and application metrics in real-time \parencite{influxdb_capitalone}. Robinhood developed their real-time risk monitoring system using InfluxDB for time series anomaly detection and Faust/Kafka for stream processing \parencite{influxdb_robinhood}.

\subsubsubsection{Data Model and Schema}

\begin{figure}[H]
\centering
\includegraphics[width=\textwidth]{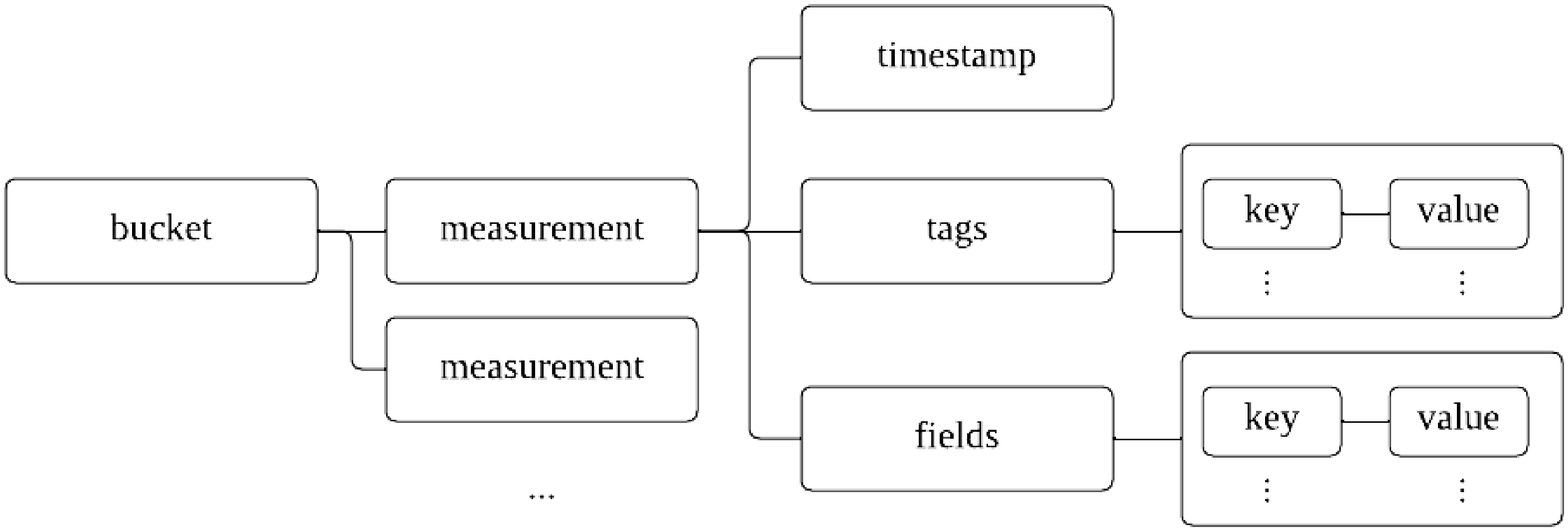} 
\caption{InfluxDB Data Model}
\label{influx model}
\end{figure}

InfluxDB has optimized its data model and schema for time-series data compared to traditional databases \parencite{naqvi2017time}. Figure \ref{influx model} demonstrates the data model in InfluxDB. Buckets are at the highest level, which is used to group data with the same retention policy. Users can specify different retention policies for different buckets and InfluxDB would automatically delete expired data. measurements inside buckets are further used to describe and provide a classification for data points. All data in InfluxDB must contain timestamps, tags, and fields. Tags are a set of key-value pairs, usually comprise of frequently queried metadata about the data point, as the tag are indexed and faster for querying. Fields are collections of key-value pairs associated with the timestamp, ideal for storing data values as they are not indexed.

\begin{figure}[H]
\centering
\includegraphics[width=\textwidth]{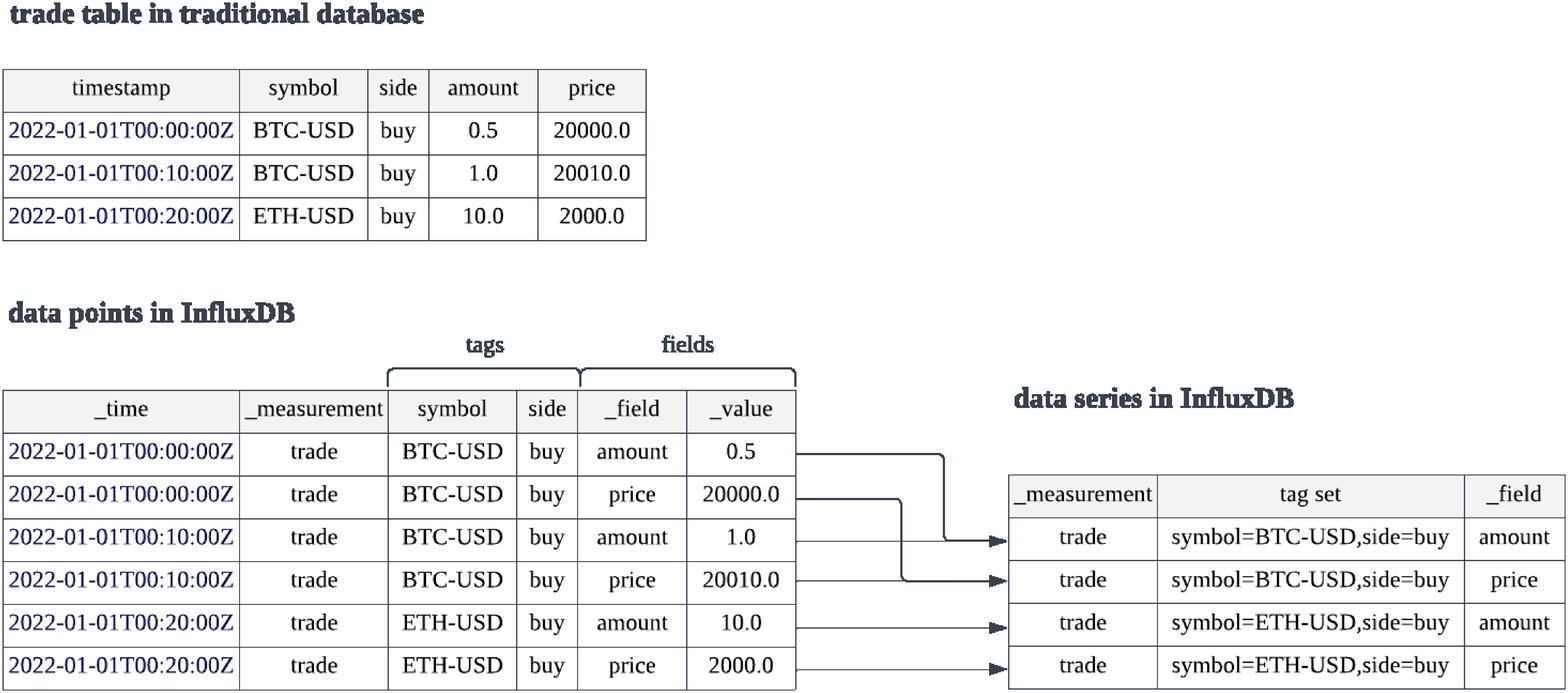} 
\caption{InfluxDB Data Schema}
\label{influx schema}
\end{figure}

Figure \ref{influx schema} shows the difference between InfluxDB schema and traditional database schema with the example of a table storing cryptocurrency trades. The trade table is described as a measurement in InfluxDB. The symbol and side columns of each trade are recorded as tags, because they are commonly used in filters and conditions in queries. The amount and price columns describe the data value of the trade and put into fields. Data series in InfluxDB contain points with the same measurement, tag set values and field key. The example data can be divided into 4 series as seen in Figure \ref{influx schema}. InfluxDB partition data according to timestamps and data series. Data points that are within the same time interval and belong to the same series are stored together in the same shard on disk. Data related to the query can be easily located using timestamp and tag values without going through data that are irrelevant to the query.

\subsubsubsection{Storage Engine}

\begin{figure}[H]
\centering
\includegraphics[width=\textwidth]{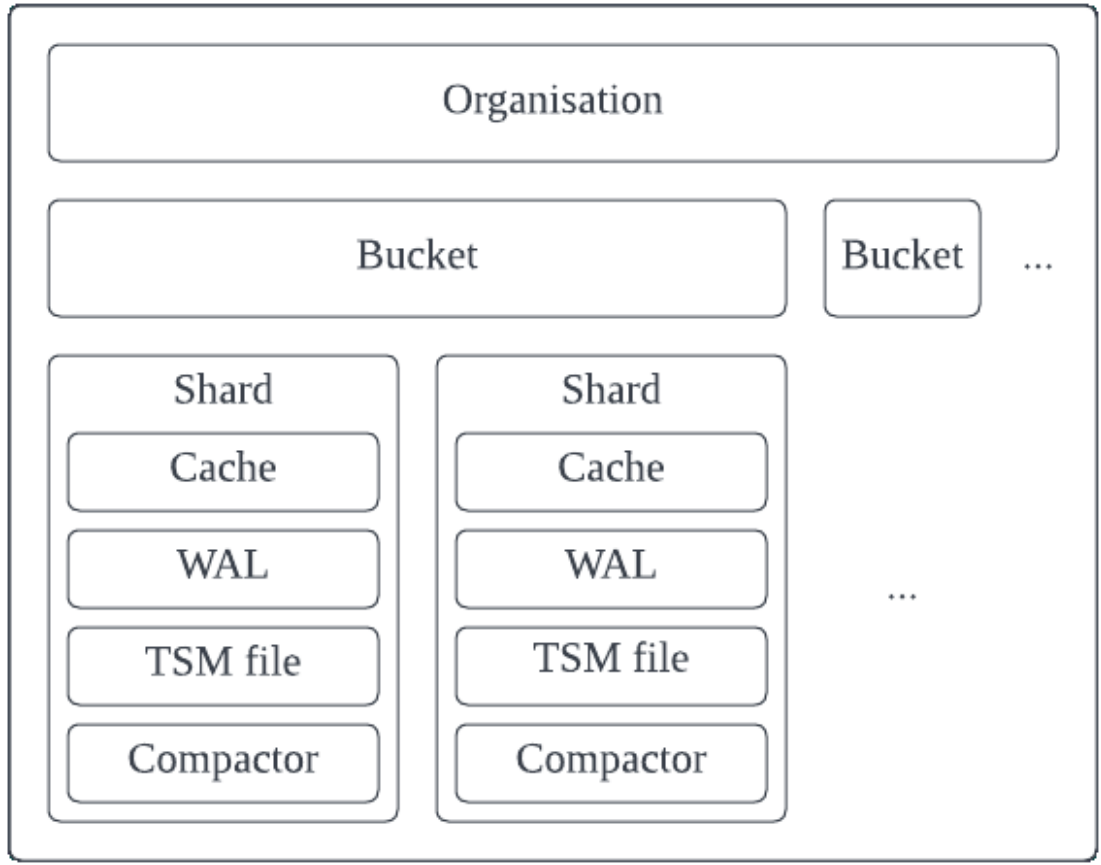} 
\caption{InfluxDB Storage Architecture}
\label{influx storage}
\end{figure}
Figure \ref{influx storage} demonstrates the storage hierarchy of InfluxDB's architecture. Data are divided into buckets with different retention policies, and further divided into shards according to timestamps. Each shard contains data series in a certain time interval. To increase write throughput, new data is written to in-memory cache and Write Ahead Log (WAL) \parencite{influxdb_TSM}. Data in cache and WAL are compacted to read-only Time-Structured Merge Tree (TSM) files on-disk to optimize for read operation. The compactor performs leveled compactions as TSM files grow large in size to reduce storage.

\begin{figure}[H]
\centering
\includegraphics[width=\textwidth]{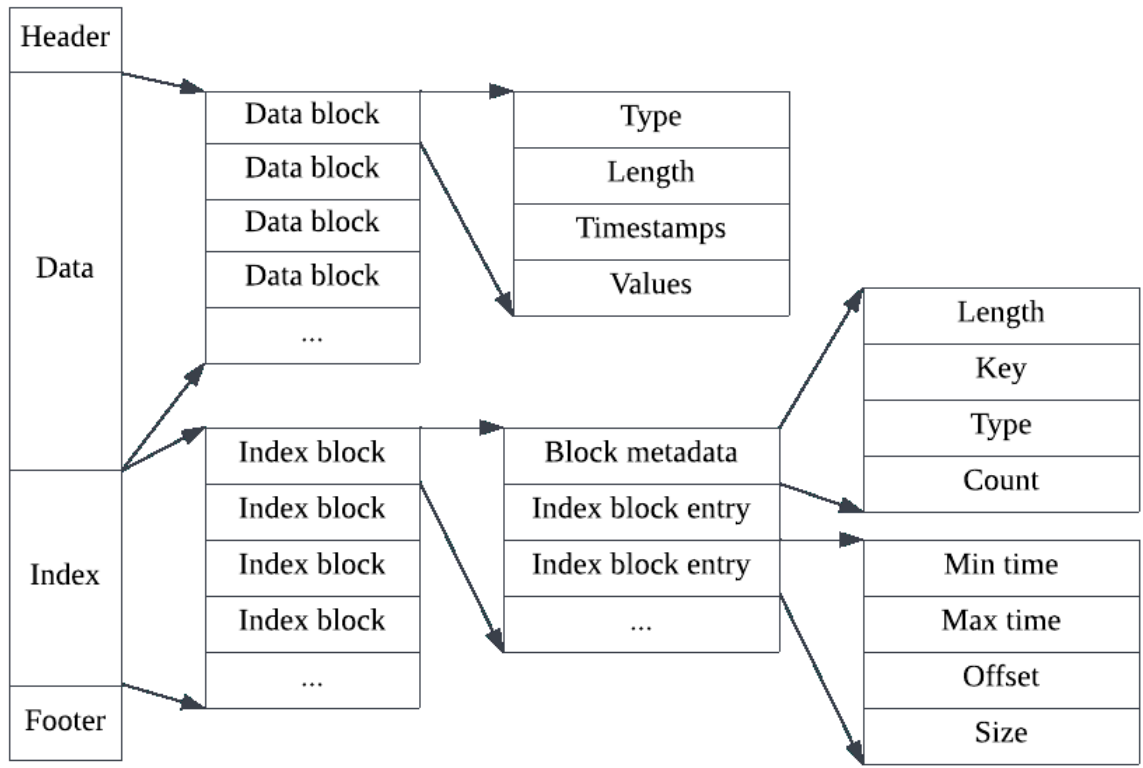} 
\caption{InfluxDB TSM File Structure}
\label{influx TSM}
\end{figure}

TSM was developed by InfluxDB to optimize for time series data, Figure \ref{influx TSM} illustrates the structure of a TSM file. A TSM file contains a data section that stores the timestamps and field values of data points and an index section that serves as the index for the data section. The data section comprises data blocks, each of which stores the data of one series in a time interval. The timestamps are compressed using delta encoding and the field values are compressed using different methods specific to the data type. A data block is the smallest unit to read when performing queries. Each of the index blocks in the index section is responsible for one series stored in the TSM file. The metadata in the index block contains information about the data series, whose key is the combination of measurement, tag set, and field key. The following index block entries contains the time interval and location of each block in the TSM file that records data for this series. The file footer is the offset of the index section. The index section blocks are ordered by the lexicographical order of the series key and the index block entries are ordered by time. This allows for fast locating of the data block using binary search given the key and timestamp.

\subsubsection{TimescaleDB}

TimescaleDB \parencite{timescale} is an open source time series database developed by Timescale Inc. It is an extension of PostgreSQL that optimises for time series data storage and processing \parencite{timescale_overview}. As an extension of PostgreSQL, it is a fully relational database. It provides high levels of analytics abilities with SQL in comparison with NoSQL databases. TimescaleDB also benefits from the PostgreSQL ecosystem of a wide array of extensions, management tools, and visualisation dashboards. TimescaleDB transforms PostgreSQL tables into hypertables and optimise query planning and execution. Hypertables can also be easily deployed in a distributed environment and store data across multiple nodes \parencite{distributed_hyptertable}. TimescaleDB also includes hyperfunctions, which is a set of functions oriented to time series data analysis operations. They extend SQL to achieve fast execution time on time series queries.

Comparing with PostgreSQL, TimescaleDB has shown to be 20x faster at inserting large scale data \parencite{timescale_postgres}. TimescaleDB can achieve 1.2x to 5x better performance in time-based queries and 450x better performance in time-ordering queries. TimescaleDB has also been compared with other time series databases including InfluxDB \parencite{influxdb}. TimescaleDB is tested to perform high cardinality inserts 3.5x faster and outperforms InfluxDB by 3.4x to 71x on complex queries.

\subsubsubsection{Hypertable}

Hypertable is the core of TimescaleDB's implementation \parencite{hyptertable}. Hypertable is a time-partitioned table where each partition is stored in a chunk. Each partition is a standard PostgreSQL table with time constraints specifying the time interval associated with this chunk, The hypertable is a parent table and the chunks are child tables in PostgreSQL. The chunk constraints allow TimescaleDB to locate the chunks that need to be read or modified when performing insert or write queries. Horizontal scaling can easily be implemented in TimescaleDB due to time-based partitioning.

A local index is built for each chunk to index the data included in the chunk. Without a global index, TimescaleDB ensures the uniqueness of primary keys by associating them with timestamps. As the data for a specific timestamp can only be stored in one chunk, the database only need to ensure the primary key value is unique in the local chunk. For recent data, the data and index of the recent chunk are stored in memory. This allows for fast insert as the time series data are usually inserted in time order and fast execution for queries on recent data. TimescaleDB employs an age-based compression strategy to adapt to historical data access patterns. Older chunks are transformed from row-oriented form to column-oriented form and compressed. The column-oriented form can deploy type-specific algorithms to compress the data.

\subsubsection{ClickHouse}

ClickHouse \parencite{clickhouse_title} is an column-oriented open source database system developed by Yandex. ClickHouse is a relational database supporting an extended version of SQL. It is built for large-scale OLAP application scenarios. Some of the properties of OLAP that ClickHouse optimises include data insert in large batches, tables containing a large number of columns, and data rarely updated after insertion \parencite{clickhouse_olap}. Additionally, ClickHouse expects queries to operate on a large number of rows and a small number of columns. As mentioned in Section~\ref{sec:kdb}, column-oriented layout which is also implemented by ClickHouse is suitable for OLAP workloads. ClickHouse also utilises vectorization when executing queries. Vectorization operates on values for a set of rows at a time instead of only operating on the value for a single row like traditional databases. Vectorized query execution can make better use of the CPU cache \parencite{clickhouse_vector}. ClickHouse has shown to be 24x faster than Druid and 55x faster than TimescaleDB through benchmarking experiments \parencite{clickhouse_bench}.

\subsubsection{MergeTree and Sparse Index}

MergeTree \parencite{clickhouse_mergetree} is the main storage engine in ClickHouse. When data is inserted into ClickHouse, it is first divided into parts. Each part is a small amount of data ordered by the primary key and each part has its own index. After new data has been written to small part files on disk, ClickHouse performs merge operations and combines them into larger parts in the background. Storing data in small part files allows for fast insertion while larger part files lead to efficient read operations.

\begin{figure}[H]
\centering
\includegraphics[width=\textwidth]{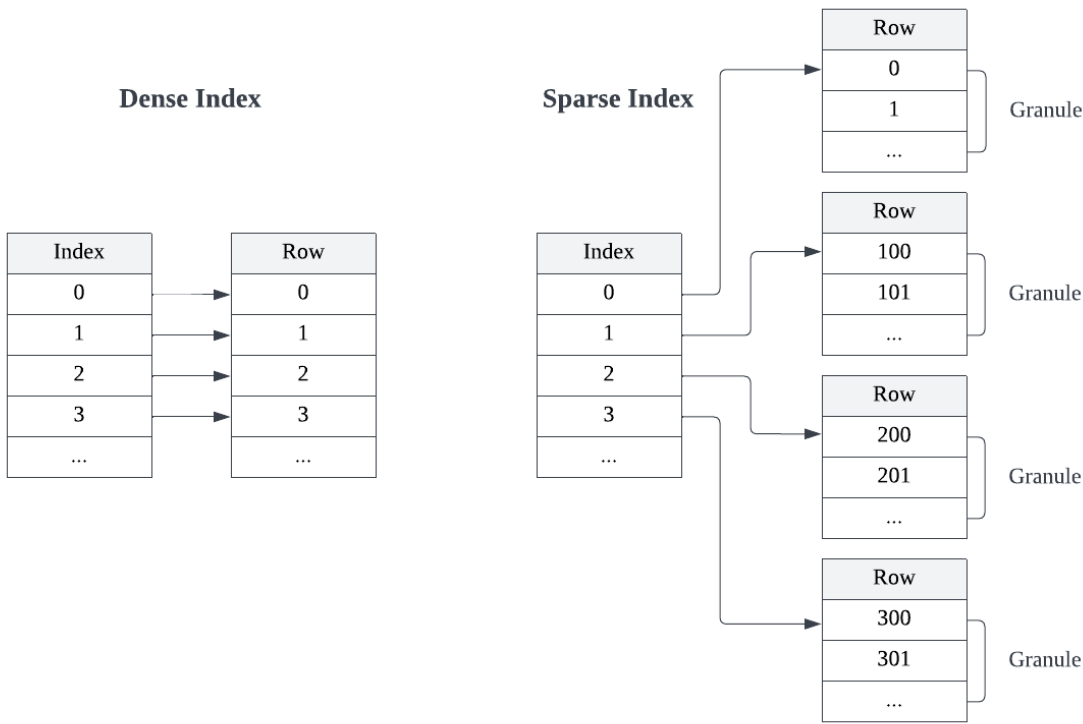} 
\caption{Dense Index and Sparse Index}
\label{sparse index}
\end{figure}

Figure \ref{sparse index} demonstrates the sparse index used by ClickHouse as the primary index. An index acts as a lookup table to locate a row quickly rather than search the whole table. Traditional databases use dense index, which means that every row in the table is recorded by an entry in the index. On the other hand, only some of the rows in the table are recorded in the sparse index. ClickHouse uses sparse index as the primary index of a table \parencite{clickhouse_sparse}. All rows in a table are divided into granules of the same size and each entry in the sparse index records the location of the first row in the granule. Granule is the smallest unit for ClickHouse to read data from disk. Using sparse index can significantly reduce the size of the index, thus allowing it to be stored in main memory for quick access. Sparse index also has good performance in range queries. A granule is the smallest unit of data that ClickHouse reads from disk and the granule size is configurable.

\subsection{Database Benchmarking}

A database benchmark is a set of experimental frameworks designed to evaluate and compare the performance of database systems \parencite{Bonnet2018}. The benchmark is defined in terms of the system under test and its execution environment for comparable results. The workload of a benchmark is a sequence of operations on the database which are generally modeled from real-world use cases and access patterns. Benchmark metrics quantify the goals of the benchmark such as throughput, latency, and storage.

Database benchmarks have the properties of \parencite{gray1992benchmark}:
\begin{itemize}
    \item Relevance: The workload and metrics simulate typical system operations.
    \item Simplicity: The workload is understandable.
    \item Portability: The benchmark can be easily reproduced on other systems rather than depending on certain functionalities.
    \item Scalability: The workload is applicable regardless of the system capacity or performance.
\end{itemize}

Database benchmarks can be divided into three categories \parencite{7990174}:
\begin{itemize}
    \item Micro Benchmarks: The benchmark evaluates the performance of specific system components including hardware and software, or program functions.
    \item End-to-End Benchmarks: The benchmark evaluates the performance of the whole system with simulated workload of typical use cases in the real world.
    \item Benchmark Suites: Benchmark suites achieve a comprehensive evaluation of the system by combining micro and end-to-end benchmarks.
\end{itemize}

Benchmarking is needed to evaluate and implement academic research on industrial solutions. Because of the large size and unique access pattern nature of time series data, the Transaction Processing Council (TPC) benchmarks are not suitable for time series databases as they are primarily designed and used with relational databases \parencite{9458659}. A number of benchmarks have been developed for time series data processing in terms of different application contexts. IoTAbench \parencite{arlitt2015iotabench} was developed for Internet-of-Things analytics scenarios with the use case of a smart metering system. It was tested against a database of 22.8 trillion readings and 727 TB of data, and it was shown to be extended to other IoT use cases. Linear Road \parencite{arasu2004linear} is a stream processing benchmark simulating a traffic monitoring system optimized to reduce traffic congestion and accidents.

\subsubsection{STAC-M3}

STAC-M3 \parencite{STAC-M3} is a benchmarking suite developed by Securities Technology Analysis Center (STAC) for tick financial analysis using time series data. The STAC Benchmarking Council consists of leading financial companies and technology providers who aim to develop and implement standardized benchmarks for various financial applications. The Antuco Suite in STAC-M3, often referred to as the baseline suite, performs a wide range of workloads on a smaller dataset sampled from real-world large dataset \parencite{STAC-M3_weka}. The Kanaga Suite, which is the scaling suite, uses a subset of Antuco queries on a dataset that's significantly larger to evaluate the performance in a scalable distributed system \parencite{STAC-M3_weka}. Each query in STAC-M3 is tested in different conditions where the number of concurrent requests and the size of the dataset vary. This is beneficial to understanding the differences between various solutions and making decisions according to the requirements of the specific use case.

However, STAC-M3 is not open source, and the specifications and dataset used for STAC-M3 are not publicly accessible. STAC-M3 evaluates the whole system stack including the database software and storage hardware. STAC-M3 has mostly been tested with kdb+, with a small number of reports on eXtremeDB and Shakti.

\subsubsection{YCSB}

The Yahoo! Cloud Serving Benchmarking (YCSB) \parencite{cooper2010benchmarking} is an open-source benchmarking suite designed for large-scale NoSQL applications. Any benchmark in YCSB is a combination of client, workload generator, and core. The client connects to the database and executes queries. The workload generator is designed to support various workloads that simulate different application scenarios. The core package is responsible for testing the workload and recording performance metrics. In combination with the configurable workload, YCSB focuses on modeling various use cases using standard operations rather than specialized features. This makes YCSB easily applicable to diverse use cases on a wide range of databases.

YCSB has been used to compare the performance of Relational Database and NoSQL Database with MySQL and MongoDB \parencite{pandey2020performance}. It has also been used to comparatively analyse popular NoSQL databases including Redis, MongoDB, and Cassandra \parencite{9585956}. As a continuously developing open source project, YCSB supports 5 Relational Databases and 40 NoSQL databases \parencite{YCSB_guide}. YCSB++ \parencite{patil2011ycsb++} is a set of extensions to YCSB with consistency measurements and multi-phrase workload for advanced optimization features and complex requests.

%%%%%%%%%%%%%%%%%%%%%%%%%%%%%%%%%%%%
\section{Benchmark Design}
This project aims to design a benchmarking suite targeting the application of financial data analysis to assess the performance of time series databases. Firstly, a scenario where such analysis is needed was defined, along with the potential use cases. Then, related works under such use cases are investigated in terms of the specific analysis used. The benchmarks are developed according to measurements and calculations that are commonly used in real-world analysis and can serve as the basis of more complex analysis. The benchmarks also include frequently performed database actions. This chapter details the definition of the use cases and describes each benchmark in the designed suite.

\subsection{Use Cases}\label{sec:use cases}

This project defines a scenario where a database system is needed for storing and querying data from markets such as stock exchanges and cryptocurrency exchanges. The database stores two collections of data, trade data and order book data. The order book describes all active orders currently placed on a financial instrument. Executed orders are moved into transactions in the trades data. These data are time series data as each data point is associated with a timestamp and the value change over time. The data could be collected for various financial instruments traded in an exchange. It could be from multiple exchanges trading the same financial instrument as well. 

Traders could use such a system to make real-time analysis to help with their decision-making. Researchers and scholars can gain insights into the market by investigating historical data over a period of time.
Table \ref{tab:use case} concludes the use cases for such database systems in the categories of writing data, reading and querying data, and fault tolerance and efficiency of the system. These categories include different components and functionality that are related to the performance of the database system.

\begin{table}[H]
    \centering
    \begin{tabular}{ |m{3.0cm}|m{11.0cm}| } 
    \hline
\textbf{Category} & \textbf{Use Case} \\
\hline

\multirow{3}{=}{Writing} 
& The system should be able to parse and congest data from exchanges in real time \\\cline{2-2}
& The system should be able to perform bulk insert of a relatively large amount of data \\\cline{2-2}
\hline

\multirow{5}{=}{Reading and Querying} 
& The system should be able to make real time data available for querying \\\cline{2-2}
& The system should be able to perform relatively simpler analysis on intraday data \\\cline{2-2}
& The system should be able to perform relatively complex analysis on large amount of historical data \\\cline{2-2}
\hline

\multirow{3}{=}{Fault Tolerance and Efficiency} 
& The system should be able to store data in persistent storage and be able to recover from failures \\\cline{2-2}
& The system should be able to store the data in a memory efficient way\\\cline{2-2}
\hline

\end{tabular}
\caption{Categorised Use Cases for Database System}
\label{tab:use case}
\end{table}

\subsection{Benchmarks}

Academic studies and research related to data analysis using trades and order book have been investigated to develop the benchmarking suite. The benchmarks are developed aiming to include the most commonly used analysis and various aspects of the database system. Table \ref{tab:benchmarks} shows the specifications of each benchmark. Each benchmark is given an ID based on its content for ease of reference. The benchmarks are designed as queries that simulate real analytical queries from the users of the system. 

\begin{table}[H]
    \centering
    \begin{tabular}{ |m{1.7cm}|m{1.6cm}|m{7.0cm}|m{1.6cm}|m{1.6cm}| } 
    \hline
\textbf{Category} & \textbf{ID} & \textbf{Specification} & \textbf{I/O \newline Intensity} &\textbf{Compu-tation \newline Intensity} \\
\hline

\multirow{5}{=}{Trades} 
& T-V1 & Average trading volume over 1 min intervals in a day & light read & light \\\cline{2-5}
& T-V2 & Average daily trading volume over a month & heavy read & light\\\cline{2-5}
& T-VWAP & Volume Weighted Average Price (VWAP) over 1 min intervals in a day & light read & heavy\\\cline{2-5}
\hline

\multirow{14}{=}{Order Book} 
& O-T & Produce the top of the book at a random time & light read & light \\\cline{2-5}
& O-B1 & Find the highest bid price in a week & light read & light\\\cline{2-5}
& O-B2 & Find the highest bid price in a month & heavy read & light\\\cline{2-5}
& O-S & Bid/ask spread over a day & light read & heavy\\\cline{2-5}
& O-V1 & Find the Market Depth at the top level using the average volume with 1 minute intervals over a week & light read & heavy\\\cline{2-5}
& O-V2 & Find the Market Depth at the 5th level using the average volume with 1 hour intervals over a month & heavy read & heavy\\\cline{2-5}
& O-NBBO & National Best Bid and Offer (NBBO) for a day & light read & light\\\cline{2-5}
\hline

\multirow{10}{=}{Complex Query} 
& C-R & Mid-quote returns for a day in 5 min intervals & light read & light \\\cline{2-5}
& C-VT & Compute the volatility as the hourly standard deviation of execution price returns in 5 min intervals for a day & light read & heavy \\\cline{2-5}
& C-VO1 & Compute the volatility as hourly the standard deviation of mid-quote returns in 5 min intervals for a day & light read & heavy \\\cline{2-5}
& C-VO2 & Compute the volatility as 4-hourly the standard deviation of mid-quote returns in 1 hour intervals for a week & light read & heavy \\\cline{2-5}
\hline

Writing & W & Write one day trades and order book data to persistent storage & heavy write & light \\\cline{2-5}
\hline

Storage \newline Efficiency & SE & Compare disk storage space of same dataset in database and in raw data files & - & - \\\cline{2-5}
\hline

\end{tabular}
\caption{Benchmark Specifications}
\label{tab:benchmarks}
\end{table}

The benchmarks are divided into 5 categories. The trades category includes calculations on trades data commonly used in the industry such as average trading volume and Volume Weighted Average Price (VWAP). In addition, these calculations are relatively easy to perform without complex queries that include joins. The second category involves actions and analysis of the order book data that are frequently used such as bid/ask spread and National Best Bid and Offer (NBBO). Benchmarks that would require a relatively complicated query with joins and nested queries are classified in the complex query category. The calculations performed in these queries such as the mid-quote returns are commonly used by researchers and often serve as the basis of more complex custom analysis. Benchmark with ID W was developed to assess the bulk writing ability of the database system. The databases are also evaluated by their storage efficiency.

Each benchmark is also classified based on its Input/Output (I/O) intensity and computation intensity. I/O intensity describes the amount of data that needs to be read or written during execution. I/O intensity is heavy when the data required to access for the query span across a large number of data files or blocks. Computation intensity categorizes the queries in terms of the amount of computation performed on the data in order to execute the query. Queries with heavy computation intensity don't necessarily require a large dataset but use the data for complex calculations. The benchmarks vary in terms of their classification in I/O intensity and computation intensity to cover different aspects of the database system.

\subsubsection{Trades}
\subsubsubsection{Trade Volume}

\begin{equation}
\begin{gathered}
Average Volume = \sum_{t \in T} t.amount \\
where \: T = \{t \: | \: t.timestamp \in t_i, t.side = s_i, t.symbol = sym_i\}
\label{eq:volume}
\end{gathered}
\end{equation}

Equation \eqref{eq:volume} shows the formula for calculating average volume. \(t_i\) is an element from a set of time intervals. \(s_i\) is from \(\{buy, sell\}\) which indicates whether the trade was executed on the buy side or sell side. \(sym_i\) are symbols of instruments being traded that are recorded in the dataset. The average volume is calculated by the sum of the trading amount of all orders that are within a certain time interval, executed on one side, and trade a specific symbol. Trade volume has been recognised as a traditional measure of market liquidity analysis \parencite{muranaga1999market}. Average trading volume can be used to develop trading strategies as it has been found to be strongly related to stock returns and predictive of future returns \parencite{chandrapala2011relationship}. 

To implement benchmark T-V1, \(t_i\) is an element of a set of 1-minute intervals over a specific day. \(sym_i\) are all symbols recorded during this day. T-V2 is queried by setting \(t_i\) as 1-hour intervals in a 1-month period.

\subsubsubsection{Trade Price}

\begin{equation}
\begin{gathered}
VWAP =  \frac{\sum_{t \in T} (t.amount \times t.price) }{\sum_{t \in T} t.amount } \\
where \: T = \{t \: | \: t.timestamp \in t_i, t.symbol = sym\}
\label{eq:vwap}
\end{gathered}
\end{equation}

Volume Weighted Average Price (VWAP) is calculated with Equation \eqref{eq:vwap}. \(t.amount \times t.price\) represents the total value of the traded asset, while VWAP is calculated using the total traded value divided by the total traded volume. VWAP provides more insights than simply the trading price as it is adjusted to the volume. VWAP can be used as a benchmark price to assess the market impact of a trade execution \parencite{berkowitz1988total}. Trading strategies can be developed to optimize for VWAP. To calculate benchmark T-VWAP, \(t_i\) is every element in a set of 1-minute intervals over a specific day. \(sym\) is the symbol of the specific asset being traded which is targeted by the calculation.

\subsubsection{Order Book}

\begin{figure}[H]
\centering
\includegraphics[width=0.7\textwidth]{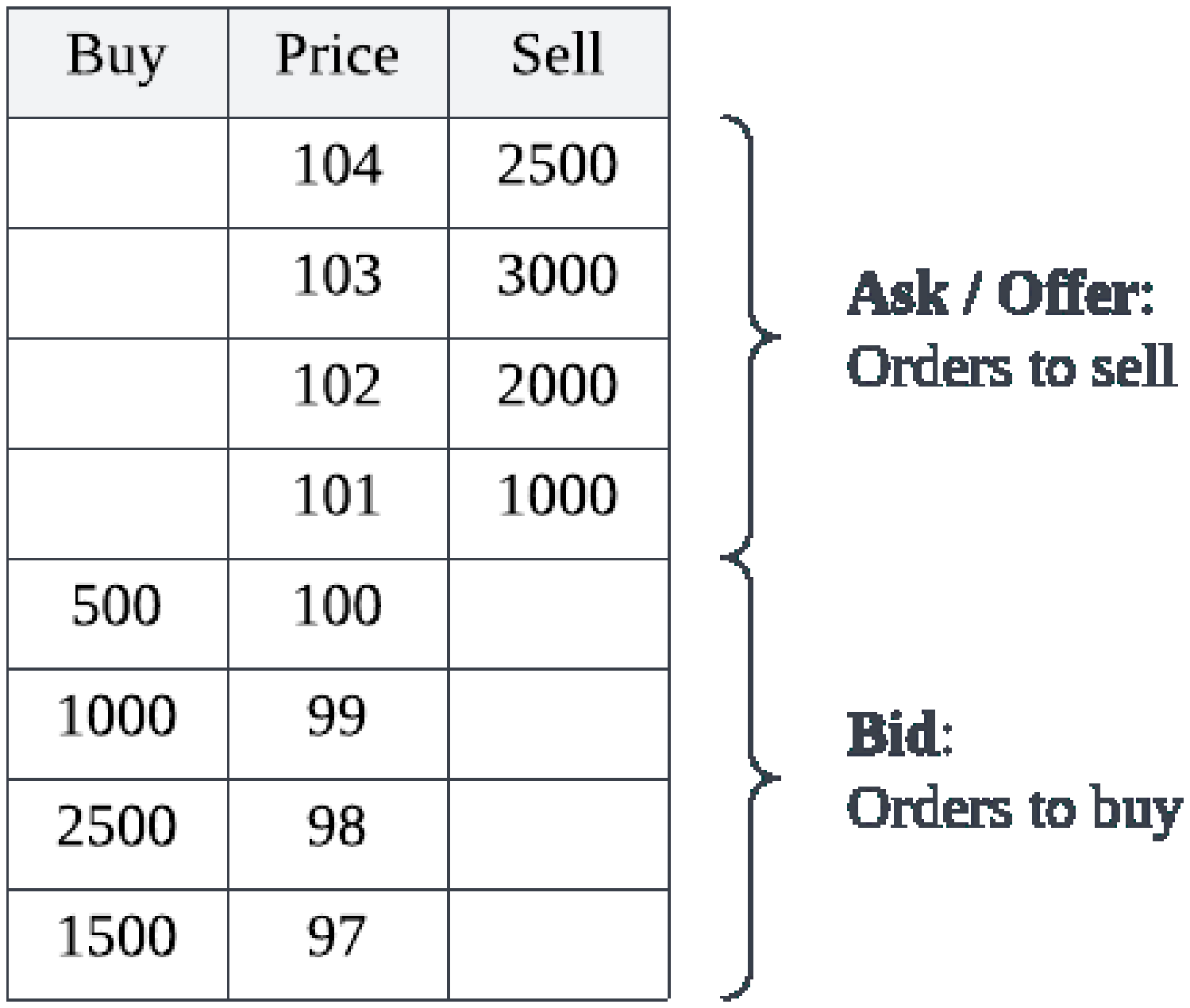} 
\caption{Order Book Example}
\label{order book}
\end{figure}

Figure \ref{order book} illustrates an example of an order book. An order book is an organised list of active buy and sell orders placed on an instrument. The sell side is often referred to as ask or offer, while the buy side is called bid. The order book is organised by the price level, which is the price at which multiple orders wish to buy or sell. Each order contains the buyer or seller's information, the amount, and the price they want to execute. As an example, the first row of Figure \ref{order book} shows that all sell orders placed at the price of 104 accumulate to 2500 units. In electronic exchanges, a matching engine is used to match buy and sell orders at the same price level to execute the trade. Executed orders will become trades and be removed from the order book so the ask and bid sides do not cross over.

\subsubsubsection{Top of the Book}

Top of the Book is the highest bid price and the lowest ask price of the order book. In the example in Figure \ref{order book}, the Top of the Book is price levels 100 and 101. The Top of the Book shows the requirement for an order to be fulfilled. It also has an effect on order placing strategies and activities including order submission, cancellation, and amendment \parencite{cao2008order}. Benchmark O-T chooses a random timestamp in the dataset and asks for the current Top of the Book for a specific symbol. This is an operation that is expected to be performed fast during intraday trading time as it requires relatively very little disk read and computation.

\subsubsubsection{Order Price}

Related to the Top of the Book, O-B1 and O-B2 find the highest bid price in a certain time interval. This operation could be performed when doing an analysis of historical data or making a comparison of current market data with historical data. O-B1 and O-B2 require light computation intensity as it is a simple aggregation operation to find the highest value for a series of data. These benchmarks differ in the time interval they operate on, which reflects the amount of data needs to be read. They are designed to assess the scalability of the database system with respect to the amount of data.
\begin{equation}
\begin{gathered}
Bid-Ask \: Spread =  MIN(ask) - MAX(bid)
\label{eq:bid ask spread}
\end{gathered}
\end{equation}
Equation \eqref{eq:bid ask spread} demonstrates that the Bid-Ask Spread is calculated by the lowest ask price deducting the highest bid price. Bid-Ask Spread is the minimum risk or cost of a transaction. It is often used as an indicator for market liquidity as assets with a significant amount of liquidity tend to have a small spread. It is also used for the analysis of intraday behavior of securities \parencite{abhyankar1997bid}, which is the use case simulated by benchmark O-S.

National Best Bid and Offer (NBBO) is the highest bid price and lowest ask price for a financial instrument that is calculated across multiple exchanges. NBBO benefits traders by finding the best prices in different exchanges. Benchmark O-NBBO is designed to use for intraday analysis and to increase the volume of the dataset from a single exchange to multiple exchanges.

\subsubsubsection{Order Volume}

\begin{equation}
\begin{gathered}
Market \: Depth = \sum_{o \in O} o.amount \\
where \: O = \{o \: | \: o.price \in \{l_1, l_2 ... l_n\}, o.side = s_i, o.symbol = sym\}
\label{eq:market depth}
\end{gathered}
\end{equation}

The volume of assets willing to be traded for each order is included in the order book. The equation \eqref{eq:market depth} shows the calculation of Market Depth at level \(n\) using the order volume. The Market Depth at any time is the accumulated order volume of all orders on the buy or sell side whose prices do not exceed the specified price level. As an example, the Market Depth at price level \(n\) on the sell side is the sum of order volume at all price levels whose price is less than the price of level \(n\). Market Depth is the amount of asset that must be traded in order to move the price to price level \(n\). Market Depth is an indicator of market liquidity. Market liquidity is high when Market Depth is deep, which means that the asset price is likely to remain stable when trades are executed. Market Depth has been found to have an impact on price discovery and the volume imbalance which can be calculated from Market Depth is related to short-term returns in the future \parencite{cao2009information}.

To calculate the Market Depth for a given time period, the time period is split into time intervals where the order volume for each interval is the average order volume of all order book snapshots during the time interval. Benchmark O-V1 and O-V2 use the time period of 1 week and 1 month, the time interval of 1 minute and 1 hour respectively.

\subsubsection{Complex Query}

\begin{equation}
\begin{gathered}
Return = log(price_t) - log(price_{t-1})
\label{eq:return}
\end{gathered}
\end{equation}

\begin{equation}
\begin{gathered}
Execution \: Price \: Return = Return \: where \: price = t.price
\label{eq:execution return}
\end{gathered}
\end{equation}

\begin{equation}
\begin{gathered}
 Mid\: Quote \: Return = Return \: where \:  price = \frac{MIN(o.ask) - MAX(o.bid)}{2}
\label{eq:mid quote return}
\end{gathered}
\end{equation}
The return for a given asset at time \(t\) is calculated as the logarithmic difference of the price between time \(t\) and time \(t-1\), as shown in Equation \eqref{eq:return}. The execution price return in Equation \eqref{eq:execution return} is calculated using the price of the executed trade at time \(t\) and \(t-1\). Mid quote price, which is the average of the lowest ask price and highest bid price, is used in Equation \eqref{eq:mid quote return} to calculate mid quote return. To find the returns over a given time period, the time period is divided into multiple time intervals. The closing price, which is the price data with the latest timestamp in the time interval, is extracted into a list. The returns are thus calculated by applying Equation \eqref{eq:return} on every element of the list.

\begin{equation}
\begin{gathered}
 Volatility = Standard \: Deviation (R)\\
 where \: R = \{return | return.timestamp \in t_i\}
\label{eq:volatility}
\end{gathered}
\end{equation}
Equation \eqref{eq:volatility} demonstrates the analysis of market volatility using the return. Volatility is calculated by the standard deviation of the returns over a certain time period. Volatility refers to the degree of uncertainty that the price of an asset changes. A higher volatility means that the price change can be more drastic or unpredictable rather than a steady increase or decrease. Volatility help traders make a decision by indicating the short-term risk \parencite{muranaga1999market}.

Benchmark C-R calculates mid quote returns in 5-minute intervals for intraday analysis. Benchmark C-VT and C-VO1 compute the volatility using the 5-minute price returns accumulated over each hour. Benchmark C-VO2 calculates the volatility for a longer period of time using data with lower granularity, which is designed to simulate historical data analysis. 

\begin{lstlisting}[language=SQL, caption={Impelementation of C-VO1 in PostgreSQL}, label = {postgres example}]
SELECT time_bucket('1 hour', timestamp) AS one_hour, 
       stddev(return) AS volatility 
FROM
(
    SELECT timestamp, 
           LOG(a1price/2+b1price/2) - LOG(LAG(a1price/2+b1price/2) OVER ()) AS return 
    FROM
    (
        SELECT time_bucket('5 minutes', timestamp) AS five_min, 
               max(timestamp)
        FROM books
        WHERE timestamp > TIMESTAMP '2022-06-18 00:00:00' 
              AND timestamp < TIMESTAMP '2022-06-19 00:00:00' 
              AND sym = 'BTC-USD'
        GROUP BY five_min
    ) AS last_books
    INNER JOIN books
        ON books.timestamp = last_books.max
) AS log_diffs
GROUP BY one_hour
;
\end{lstlisting}

The benchmarks in the complex query category are designed to assess database systems on their ability to execute complex queries including nested queries and joins. Listing \ref{postgres example} is an example implementation of C-VO1 in PostgreSQL with two nested queries and one inner join. This query computes the volatility for a day using the mid quote returns. The innermost query finds the timestamps of the records with the closing price in each 5-minute interval. This is joined with the whole order book table to produce a table of these records. The mid-quote returns are then calculated as the difference between the logarithm of mid-quote prices of adjacent rows. The \verb|LAG| function is used to access the previous row. The outermost query computes the volatility as the standard deviation of returns in 1-hour intervals.

\subsubsection{Writing}

Benchmark W measures the bulk insert performance of database systems. The data is inserted into the database from two files in a common file format such as \verb|.csv|. The same files, storing one-day trades and order book data respectively, are ingested into persistent storage. This benchmark records the time between starting to load and parse the file until the data is in the database storage and ready to be queried. The schema of the data in the file is provided in the event that it is required by the database system.

\subsubsection{Storage Efficiency}

\begin{equation}
\begin{gathered}
Storage \: Efficiency = \frac{Size\: of\: data\: in\: database\: storage}{Size\: of\: data\: on\: file} \times 100\%
\label{eq:storage}
\end{gathered}
\end{equation}

Benchmark SE evaluates the storage efficiency with Equation \eqref{eq:storage} expressed in percentage. Database systems often utilize compression and compaction techniques to reduce the disk storage size of the database. However, more compression and compaction also increase the time needed to unpack the data when executing queries. A trade-off exists between storage efficiency and read performance. Query performance can be improved by indexing, which is deployed by many database systems to locate data fast. On the other hand, indexing could take up extra disk space besides the actual data values and the indexing file size grows with the database size.

\subsection{Metric}

The main metric measured for each of the benchmarks the query latency. Query latency is the amount of time between sending the query to the server and receiving the results of the query. Query latency includes 2 parts, the time spent sending data through the network and the query execution time on the server side. However, network latency cannot indicate the database performance and it is dependent on aspects such as network bandwidth and physical distance. Thus it is assumed that the user has a reliable network connection with the server such that the network latency can be ignored. As the network latency is ignored, the query execution time can be obtained as the query latency using the built-in timer of each database system.
%The total amount of space used during query execution is also collected from kdb+ and InfluxDB using their built in functionality.

The secondary metric is the total amount of storage space used during query execution. Although efficient use of storage is desirable for a database, query latency is more important when responding to client queries. As storage devices get cheaper, limitations in storage can be easily overcome by adding horizontal storage.

%%%%%%%%%%%%%%%%%%%%%%%%%%%%%%%%%%%%
\section{Data Collection}
This study uses data collected from cryptocurrency exchanges as an example of the data collected in real-world use cases as defined in Section \ref{sec:use cases}. The data used to implement the benchmarking suite includes two kinds of data. The real trading data from cryptocurrency exchanges in real-time is collected using a data collection pipeline. Additionally, random data is generated to simulate real-world data. This chapter details the data collection process and presents the dataset.

Cryptocurrency exchanges are online platforms where people can trade assets such as cryptocurrencies and digital currencies. They perform the same role as conventional stock exchanges in the market. As cryptocurrency exchanges are implemented with blockchain technology, the activities performed on these exchanges are visible to everyone on the network. Real-time trading data and historical data are easily accessible to users. Due to the distributed nature of the block-chain, the data is also reliable as it has been validated by the miners. The trades in cryptocurrency exchanges are identical to trades in real-world exchanges besides the difference in the currency. Depending on the specific exchange, a cryptocurrency exchange could have a large volume of data stream that is similar to a real-world exchange. These reasons make cryptocurrency exchange data ideal for simulating the data and workload of real-world financial applications.

\subsection{Data Collection Pipeline}

Cryptofeed \parencite{cryptofeed} is used for collecting data from cryptocurrency exchanges in this project. It is an open-source Python library designed for handling data feed from cryptocurrency exchanges. It can be used to retrieve real-time or historical data. Cryptofeed combines the feeds from an arbitrary combination of the supported exchanges into one standardized and normalized feed. Developers can register callbacks on different feed channels including trades, ticker updates, and order book updates to handle the data.

\begin{figure}[H]
\centering
\includegraphics[width=1.03\columnwidth]{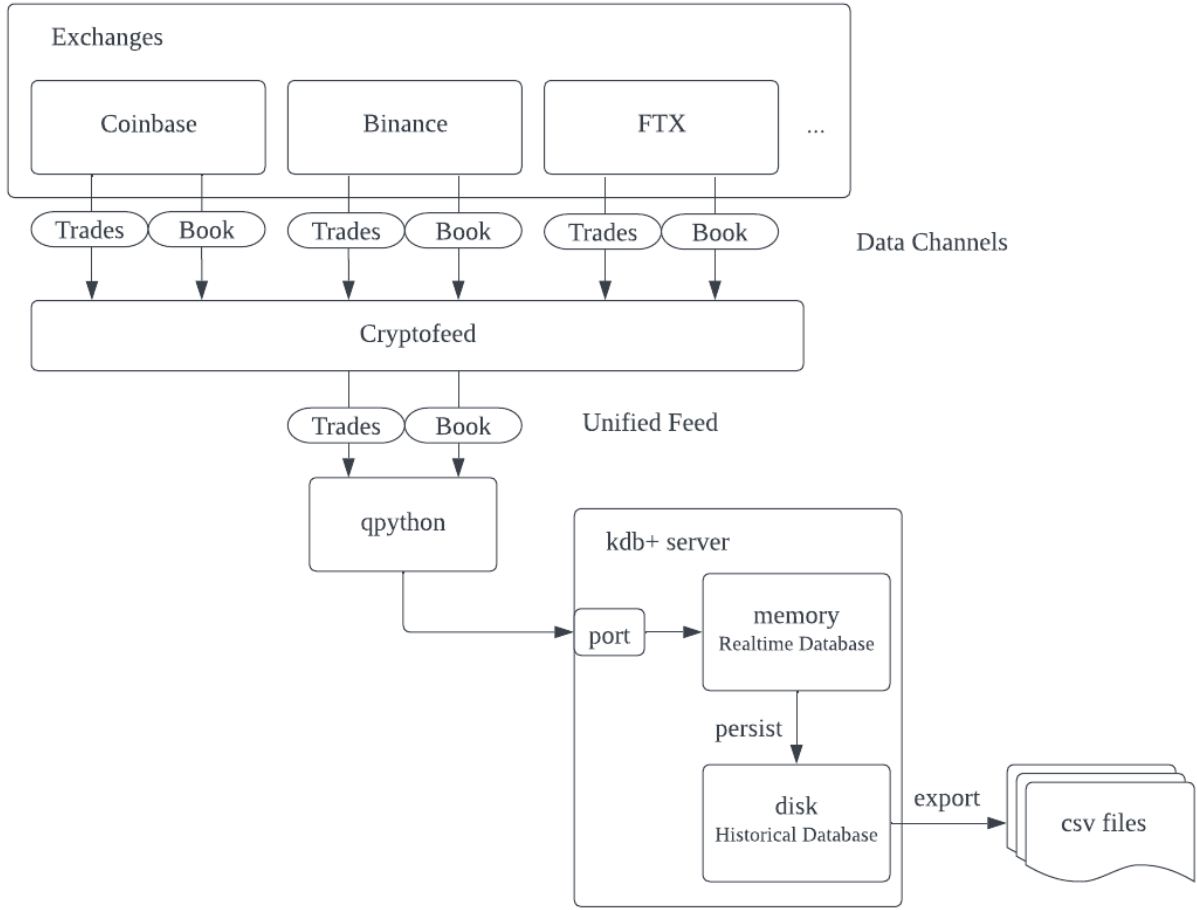} 
\caption{Kdb+ Data Collection Pipeline}
\label{kdb_collection_pipeline}
\end{figure}

Figure \ref{kdb_collection_pipeline} demonstrates the data collection pipeline using kdb+. Multiple exchanges generate their own data feeds from their operations. A FeedHandler is defined in Cryptofeed to subscribe to feeds from specified exchanges and combine them into one standardized feed where the data is transformed into Python objects. Each exchange may open multiple data channels for different kinds of data streams. This project subscribes to the trades and order book channels to collect data, and Cryptofeed generates a unified feed for each channel.

The callbacks for the unified feed generated by Cryptofeed are implemented with qPython. qPython \parencite{qPython} is a python library that enables the communication between Python and kdb+/q processes. qPython first opens on a connection with the kdb+ server, which has a q process listening on a specific port. When the data is received from the subscribed feed, Cryptofeed invokes the callback function. The callback function would preprocess the data, structure the query and send the data to the kdb+ server for storage. A qPython callback is implemented for each channel.

When the kdb+ server receives the data on the open port, it first saves the data in the Realtime Database in memory. At the end of every day, the content from the Realtime Database is persisted in the Historical Database on disk. Thus the Realtime Database can free up space to store incoming new data. This project implements the kdb+ server on the local machine and stores the trade and order book channel data in two tables respectively. The Historical Database is exported into CSV files which serve as the raw data files to import into other database systems used for testing. A separate file is generated for each table every day.

\subsection{Random Data Generation}\label{sec:random data}

One day of trades and order book data is generated randomly to implement the writing benchmark with ID W. They are generated using kdb+ and saved into two separate CSV files. 1,000,000 rows are generated for trades and 1,500,000 rows are generated for order book updates, which are in the same format as the collected real data. The generated data has the same symbols, the price and amount of each record are randomly generated according to the average value of the corresponding symbol in the collected real-world data. 

\subsection{Dataset}

As a sample of the data feeds from cryptocurrency exchanges in real-world use cases, this project collected the data of 3 symbols, namely BTC-USD, ETH-USD and USDT-USD. Real-time trades of BTC, ETH, and USDT are stored in the trades table. The order book table subscribes to level 2 order book data, which consists of the sizes of the bid and ask prices on every price level. Only the top 20 price level data are collected as they are more related to the price of the assets and more likely to be matched to execute a trade. The order book table stores a \(\frac{1}{40}\) sample of the order book updates received from the exchange. This is because order book updates occur very frequently, which means that there is a lot of redundant data if all the updates are being recorded.

The dataset can be divided into three date intervals, 2022-06-01, 2022-06-18 to 2022-07-17, and 2022-07-20. Data with the date 2022-06-01 is randomly generated as described in Section \ref{sec:random data}. Data from 2022-06-18 to 2022-07-17 is collected from only the Coinbase exchange to implement most of the benchmarks. Data collected on 2022-07-20 is only the order book from multiple exchanges including Coinbase, BinanceUS, FTX, FTXUS and Bitstamp to implement benchmark O-NBBO.

The total dataset has around 20 million records in the trades table and 33 million records in the order book table. The size of the exported CSV data files has a total size of 25.14 GB. Table \ref{tab:dataset} shows information about the dataset in terms of the symbol and tables. BTC-USD and ETH-USD have a similar number of trades and a large number of order book updates. Their trading frequency roughly doubled when trading in multiple exchanges. The trades and order book updates of USDT-USD are significantly less than the other two recorded symbols.

\begin{table}[H]
    \centering
    \begin{tabular}{ |m{2.3cm}|m{2.3cm}|m{2.2cm}|m{2.8cm}|m{3.6cm}| } 
    \hline
\textbf{Date} & \textbf{Symbol} & \textbf{Table} & \textbf{Rows(million)} & \textbf{Frequency(second)} \\
\hline

\multirow{6}{=}{2022-06-18 - 2022-07-17}           
            & \multirow{2}{=}{BTC-USD}
                        & Trades      & 8.81      & 6.86     \\  \cline{3-5}
            &           & Order Book  & 15.66     & 10.98     \\  \cline{2-5}
            & \multirow{2}{=}{ETH-USD}
                        & Trades      & 10.01     & 8.41    \\  \cline{3-5}
            &           & Order Book  & 14.06     & 10.01    \\  \cline{2-5}
            & \multirow{2}{=}{USDT-USD}
                        & Trades      & 0.76      & 1.80      \\  \cline{3-5}
            &           & Order Book  & 0.92      & 1.65     \\  

\hline

\multirow{3}{=}{2022-07-20}
& BTC-USD & Order Book & 1.43 & 16.79 \\ \cline{2-5}
& ETH-USD & Order Book & 1.22 & 14.42 \\ \cline{2-5}
& USDT-USD & Order Book & 0.11 & 1.89 \\ \cline{2-5}

\hline

\end{tabular}
\caption{Dataset Information}
\label{tab:dataset}
\end{table}

%%%%%%%%%%%%%%%%%%%%%%%%%%%%%%%%%%%%
\section{Implementation}
The benchmarking suite is implemented by writing queries corresponding to each of the benchmarks and executing them on different database systems. It has been implemented for 4 database systems including kdb+, InfluxDB, TimescaleDB, and ClickHouse. Additionally, kdb+ is implemented in both in-memory and on-disk modes. These databases are installed onto the local machine used for testing. This chapter details the 3 steps in the process of implementing the benchmarks, which are dataset importing, benchmark implementation, and benchmark execution.

The 4 databases, kdb+, InfluxDB, TimescaleDB, and ClickHouse are chosen because of their different characteristics and optimization for time series data. kdb+ has an excellent performance in benchmarks and is being widely used in the financial industry. kdb+ uses a columnar layout to store data on disk, as well as a lightweight query engine written in a special language q. The kdb+ /tick architecture offers the in-memory mode Realtime Database (RDB) and the on-disk mode Historical Database (HDB). InfluxDB is built as a general-purpose time series database. They have designed their own data models and storage files to optimise for time series data. TimescaleDB is an extension of PostgreSQL, which means that it is a fully relational database with SQL support. ClickHouse is designed for big data Online Analytical Processing (OLAP). It is optimised to deal with a large amount of data and complex analytical queries, which is required by the use case.

\subsubsection{Dataset Importing}

The dataset is imported into each of the database systems from the CSV files. The database schema and data types are kept consistent in all databases. For example, kdb+ allows for a list to be inserted into a table cell while other databases only allow a single element in table cells. To maintain consistency, the kdb+ database has the same data types as other databases and does not make use of this functionality. In terms of InfluxDB where the data model is different from others, the dataset schema is implemented in a way recommended by InfluxDB. Columns with categorical values and often used to filter data in queries are imported as tags. Columns with continuous numerical values and usually used for calculation are stored as fields. Additionally, the partitioning rules are set as 1 day for all databases.
%same schema, same data type, same format (e.g. list vs element)
%same partitioning rules
%

\subsubsection{Benchmark Implementation}

To implement the benchmarking suite, the benchmarks are transformed into queries to be executed in the 4 databases. Each benchmark is written into one query. A benchmark can be broken into 2 parts, data, and operation. The data is selected from the dataset on which the operation, such as numerical calculation or aggregation, is performed. Table \ref{tab:data to benchmark}  shows the correspondence between the benchmarks and their date range.

%benchmark to data table

\begin{table}[H]
    \centering
    \begin{tabular}{ |m{3.0cm}|m{3.0cm}|m{7.0cm}| } 
    \hline
\textbf{Date Range} & \textbf{Table} & \textbf{Benchmark ID} \\
\hline

\multirow{2}{=}{2022-06-18} 
& Trades & T-V1, T-VWAP, C-VT \\\cline{2-3}
& Order Book & O-S, C-R, C-VO1 \\\cline{2-3}
\hline

\multirow{2}{=}{2022-06-18 - 2022-06-24} 
& Trades &  \\\cline{2-3}
& Order Book & O-B1, O-V1, C-VO2 \\\cline{2-3}
\hline

\multirow{2}{=}{2022-06-18 - 2022-07-17} 
& Trades & T-V2 \\\cline{2-3}
& Order Book & O-T, O-B2, O-V2 \\\cline{2-3}
\hline

\multirow{1}{=}{2022-07-20} 
& Order Book & O-NBBO \\\cline{1-3}
\hline

\end{tabular}
\caption{Data Range and Corresponding Benchmarks}
\label{tab:data to benchmark}
\end{table}

The implementation of each database uses the query language specific to the database, such as q for kdb+ and Flux for InfluxDB. TimescaleDB and ClickHouse both use SQL as their query language. They have extended SQL with their own functions in areas including aggregation, arithmetic calculation, and grouping. These functions are preferred over native SQL implementations as they have been optimised by the database systems for better performance.
It is also made sure that the results of the same benchmark returned by different databases are consistent. The results should have the same columns with the same order of the rows. This may require additional operations than specified in the benchmark, such as ordering ClickHouse results because they are not ordered as the other databases, or pivoting InfluxDB columns because of the different data model.

The kdb+ database is split into in-memory mode and on-disk mode according to the kdb+ /tick architecture described in Section \ref{sec:kdb tick}. As the in-memory Realtime Database only stores intraday data, the in-memory mode implementation only includes benchmarks that operate on one-day data. All benchmarks are implemented in the on-disk kdb+ database.

%each benchmark = 1 query
%use database specific query language, use optimized functions
%same results returned (e.g. columns, column name, same row order)
%main metric = execution time, assumption
%side metric = space

\subsubsection{Benchmark Execution}

The benchmark execution process is automated with Python. Python makes a connection with the database server hosted on the local machine via a specific port. The queries are sent over to the database and executed. The results of the query are returned as Python objects. The machine used for executing the benchmarking suite is a personal PC with an Intel i7-1065G7 CPU and 16GB RAM. All databases save their data in SSD. Adequate memory space is made available during execution, only one database is being tested at the same time. The result for each benchmark is obtained by averaging the result from 10 executions. 

\begin{figure}[H]
\centering
\includegraphics[width=0.9\columnwidth]{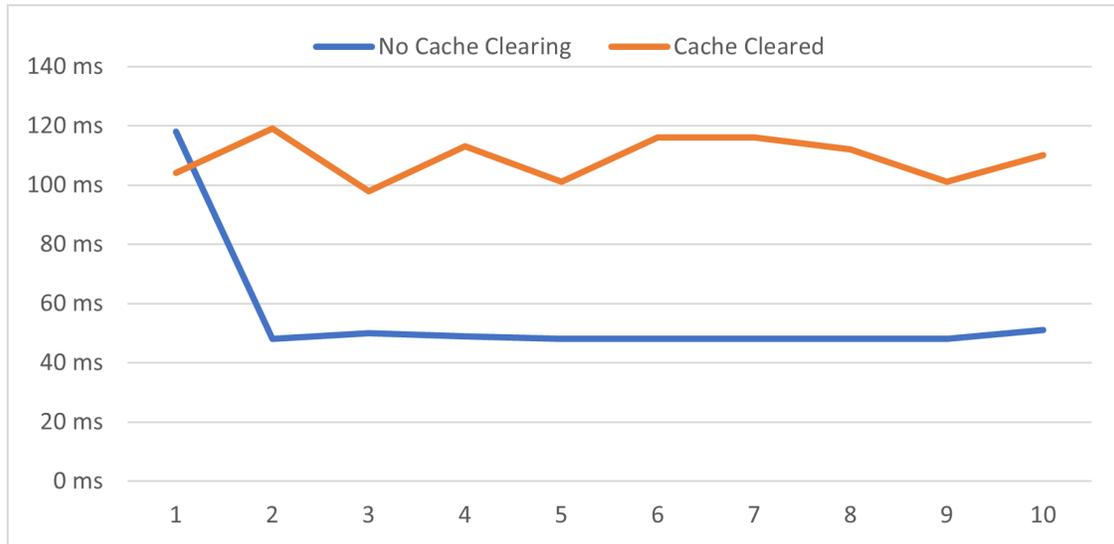} 
\caption{The Effect of Caching on Query Execution Time}
\label{fig:cache clearing}
\end{figure}

During execution, the operating system would store the data read from disk into the cache. If the same query or a similar query is executed again, the query would be executed faster as the data is read from the cache rather than disk. As database systems store data on disk, this introduces a problem where the query execution time is affected and could lead to false performance evaluation due to the cache. Figure \ref{fig:cache clearing} shows how the caching effect can influence execution time. Without clearing the cache, the execution time decrease drastically after the first execution. If the cache is cleared before every execution, the query execution time fluctuates depending on the disk access time but maintains a stable level. To avoid such issues, the standby list cache is cleared before every execution of a query.
%python
%same machine
%one database at a time
%adequate memory
%average of 10 runs
%cache clearing

%%%%%%%%%%%%%%%%%%%%%%%%%%%%%%%%%%%%
\section{Result and Evaluation}
Experimental results are obtained by executing the benchmarks in kdb+ in-memory, kdb+ on-disk, InfluxDB, TimescaleDB and ClickHouse. These results are shown in this chapter. The performances of the tested database systems are also analysed in terms of ingestion and storage, read queries, computationally intensive queries, and complex queries. Furthermore, the benchmarking suite developed in this paper is evaluated against related work.

\subsection{Ingestion and Storage}

The data ingestion and storage performance are evaluated by benchmark ID W and SE. They are not implemented in kdb+ in-memory mode as the database is kept in main memory, not persistent memory. 

\begin{figure}[H]
\centering
\includegraphics[width=0.7\columnwidth]{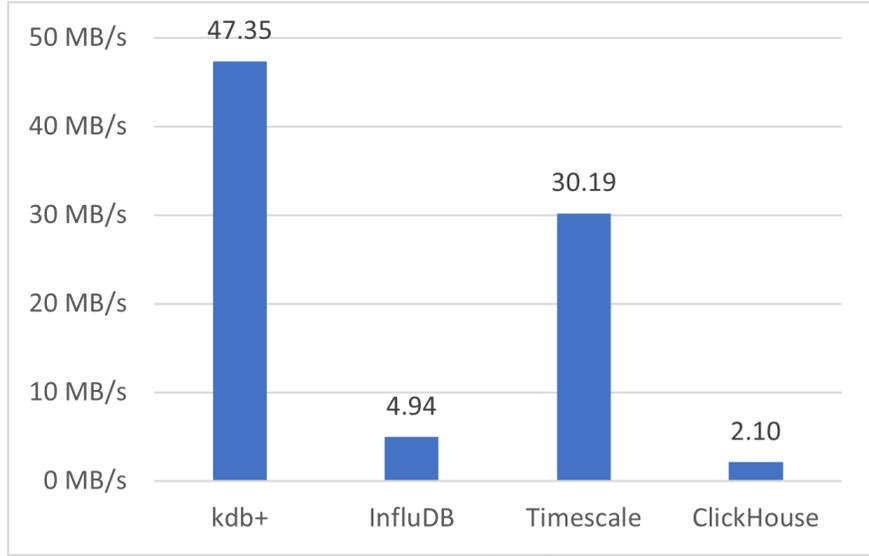} 
\caption{Data Ingestion Throughput in MB/s}
\label{fig:ingestion throughput}
\end{figure}

Table \ref{tab:load} shows the benchmarking results of each database. Benchmark W shows that kdb+ and TimescaleDB are able to import data significantly faster than InfluxDB and ClickHouse. InfluxDB is 9.6 times slower and ClickHouse is 22.6 times slower than kdb+, which is the fastest among the 4 databases. Figure \ref{fig:ingestion throughput} demonstrates the comparison of data ingestion throughput calculated from the benchmarking result. kdb+ is able to import data from CSV files at 47.35 MB/s, which is 1.6 times more than TimescaleDB. Results from benchmark SE in Table \ref{tab:load} shows InfluxDB and kdb+ has better storage efficiency than TimescaleDB and ClickHouse. InfluxDB has the best compression ratio of 83.73\%, which is 6.5\% more than kdb+ and 10.95\% more than TimescaleDB. ClickHouse's compression mechanism has relatively bad performance as the compression ratio is near 100\%.

\begin{table}[H]
    \centering
    \begin{tabular}{ |m{2.2cm}|m{2.2cm}|m{2.2cm}|m{2.2cm}|m{2.2cm}| } 
    \hline
\textbf{Benchmark ID} & \textbf{kdb+ \newline on-disk} & \textbf{InfluxDB} & \textbf{TimescaleDB} & \textbf{ClickHouse} \\
\hline
 & \multicolumn{4}{c|}{Query Latency (ms)}\\\hline

W        &  33,889 & 324,854  & 53,150   & 765,000 \\\hline

 & \multicolumn{4}{c|}{Compression Ratio (\%)}\\\hline
SE       &  90.20  & 83.73    & 94.68    & 99.65 \\\hline
  
\end{tabular}
\caption{Data Loading Performance}
\label{tab:load}
\end{table}

InfluxDB's slow data ingestion could be due to its TSM file structure and compression methods. The data schema of InfluxDB requires the data to be first classified into series to store in TSM files. Additionally, InfluxDB performs data compression before storing them in TSM files, which could be more time-consuming than other databases as InfluxDB has the best compression ratio. The index of each TSM file is built into the file by InfluxDB. As a TSM file can take up to 2GB in size, the index could be time-consuming to construct.

ClickHouse is shown to be the worst at bulk writing and storage efficiency for this project's data. This could be due to the fact that ClickHouse is designed for large-scale OLAP workloads but not specifically for time series data. This means that ClickHouse does not optimise for partitioning and ordering data according to timestamp. Additionally, ClickHouse does not use compression techniques such as delta compression for timestamps, which is commonly used by time series databases.

\subsubsubsection{Conclusion}
\begin{itemize}
    \item kdb+ has the best overall performance in this section, and it is also the best at data ingestion.
    \item InfluxDB has the best performance on storage efficiency.
    \item ClickHouse has the worst performance in this section.
\end{itemize}

\subsection{Read Queries}\label{sec:read}

Benchmarks T-V1, T-V2, O-T, O-B1, O-B2, and O-NBBO are classified as queries with mainly read operations and light computational intensity. Table \ref{tab:read} presents the results of these benchmarks. kdb+ in-memory has the best performance in terms of the benchmarks that include only one day of data. For benchmark O-NBBO, kdb+ in-memory is 7 times faster than the on-disk mode. This is expected as being in-memory significantly reduces the time spent on reading data. For the on-disk databases, kdb+ has the best overall performance with the lowest query latency for most queries. InfluxDB comes in second in the overall performance. ClickHouse is generally slower than InfluxDB while being faster than TimescaleDB.

InfluxDB is 2.7 times faster than kdb+ in T-V2 but has a similar speed in T-V1. The differences between T-V1 and T-V2 are the input data and aggregation window. T-V1 requires 1-day data while T-V2 requires 1-month data. The time window for aggregation operation is 1 minute in T-V1 and 1 day in T-V2. Since InfluxDB is significantly faster in T-V2, it could suggest that InfluxDB has better optimisation for window functions that apply aggregation to data in time windows.

\begin{table}[H]
    \centering
    \begin{tabular}{ |m{2.2cm}|m{2.2cm}|m{2.2cm}|m{2.2cm}|m{2.2cm}|m{2.2cm}| } 
    \hline
\textbf{Benchmark ID} & \textbf{kdb+ \newline in-memory} & \textbf{kdb+ \newline on-disk} & \textbf{InfluxDB} & \textbf{TimescaleDB} & \textbf{ClickHouse} \\
\hline
 & \multicolumn{5}{c|}{Query Latency (ms)}\\\hline
T-V1    & 57    &  81     & 93       & 469      & 272 \\\hline
T-V2    & -     &  740    & 273      & 3,700    & 1,991 \\\hline
O-T     & 5     &  11     & 124      & 10       & 706 \\\hline
O-B1    & -     &  94     & 146      & 3,938    & 962 \\\hline
O-B2    & -     &  287    & 440      & 12,847   & 3,202 \\\hline
O-NBBO  & 5     &  34     & 191      & 1,614    & 454 \\\hline

\end{tabular}
\caption{Performance of Read Queries}
\label{tab:read}
\end{table}

\subsubsection{Conclusion}

\begin{itemize}
    \item Read query performance: kdb+ \(>\) InfluxDB \(>\) ClickHouse \(>\) TimescaleDB.
    \item InfluxDB could be more optimised for window aggregation functions than kdb+.
\end{itemize}

\subsection{Computationally Intensive Queries}

Database performance on computationally intensive queries is assessed in benchmarks T-VWAP, O-S, O-V1, and O-V2. Table \ref{tab:comp} shows the experimental results in this category. kdb+ in-memory and on-disk modes are shown to be the fastest among all databases. There is little difference in O-S for kdb+ in-memory and on-disk. This is because most of the query latency is related to the computation time rather than the data reading time. InfluxDB has good performance in O-V1 and O-V2 while TimescaleDB and ClickHouse are significantly slower. ClickHouse has lower query latency than TimescaleDB in T-VWAP, O-V1, and O-V2 while being slower in O-S.

\begin{table}[H]
    \centering
    \begin{tabular}{ |m{2.2cm}|m{2.2cm}|m{2.2cm}|m{2.2cm}|m{2.2cm}|m{2.2cm}| } 
    \hline
\textbf{Benchmark ID} & \textbf{kdb+ \newline in-memory} & \textbf{kdb+ \newline on-disk} & \textbf{InfluxDB} & \textbf{TimescaleDB} & \textbf{ClickHouse} \\
\hline
 & \multicolumn{5}{c|}{Query Latency (ms)}\\\hline
T-VWAP  & 26    &  75     & 12,716   & 334      & 202 \\\hline
O-S     & 1,352 &  1,386  & 623,732  & 2,182    & 14,056 \\\hline
O-V1    & -     &  533    & 373      & 4,699    & 1,626 \\\hline
O-V2    & -     &  4,375  & 4,420    & 17,092   & 10,180 \\\hline
  
\end{tabular}
\caption{Performance of Computational Intensive Queries}
\label{tab:comp}
\end{table}

InfluxDB's significantly long query execution time in T-VWAP and O-S could be due to the limitation of InfluxDB's query language and data model. As InfluxDB does not offer built-in functionality for calculating the weighed average, T-VWAP is implemented using self-written reduce functions according to the equation. It could be that self-written functions are not as well optimised as built-in functions. T-VWAP and O-S both calculate their results using two field columns in InfluxDB. Two field columns in InfluxDB are separated in different series and require to be combined together using pivot before calculation. As T-VWAP and O-S do not include time window aggregation operations before the pivot operation, pivot is performed on a large number of raw data rows. The large query latency in T-VWAP and O-S could be due to the time-consuming operation of pivoting a large amount of raw data.

\subsubsection{Conclusion}

\begin{itemize}
    \item kdb+ has the best overall performance in computationally intensive queries.
    \item InfluxDB could be time-consuming when operating on multiple field columns and pivoting is required.
\end{itemize}

\subsection{Complex Queries}

Benchmarks C-R, C-VT, C-VO1, and C-VO2 evaluate database performance on complex queries. Table \ref{tab:complex} shows the obtained results respectively. Overall, kdb+ and InfluxDB have smaller query latency than TimescaleDB and ClickHouse. kdb+ and InfluxDB has similar results in C-R and C-VO1, while kdb+ is faster in C-VT and InfluxDB is faster in C-VO2. Additionally, clickhouse performs better than TimescaleDB at all benchmarks in this section.

InfluxDB is 2.8 times faster than kdb+ in C-VO2 with similar performance in C-VO1. C-VO1 and C-VO2 differ in data range and aggregation time window. This further suggests the conclusion made in Section \ref{sec:read} that InfluxDB has better performance in time window aggregation functions.

\begin{table}[H]
    \centering
    \begin{tabular}{ |m{2.2cm}|m{2.2cm}|m{2.2cm}|m{2.2cm}|m{2.2cm}|m{2.2cm}| } 
    \hline
\textbf{Benchmark ID} & \textbf{kdb+ \newline in-memory} & \textbf{kdb+ \newline on-disk} & \textbf{InfluxDB} & \textbf{TimescaleDB} & \textbf{ClickHouse} \\
\hline
 & \multicolumn{5}{c|}{Query Latency (ms)}\\\hline
C-R     & 64    &  113    & 99       & 1,614    & 401 \\\hline
C-VT    & 41    &  51     & 2,009    & 324      & 190 \\\hline
C-VO1   & 61    &  96     & 94       & 1,591    & 407 \\\hline
C-VO2   & -     &  688    & 242      & 4,319    & 1,139 \\\hline

\end{tabular}
\caption{Performance of Complex Queries}
\label{tab:complex}
\end{table}

\subsubsubsection{Conclusion}

\begin{itemize}
    \item Complex query performance: kdb+ \(=\) InfluxDB \(>\) ClickHouse \(>\) TimescaleDB.
    \item InfluxDB performs well in time window aggregation functions.
\end{itemize}

\subsection{Query Storage}

Due to the limitations of built-in functionalities for the databases, the storage space used in query execution is only measured in kdb+ and InfluxDB. Table \ref{tab:storage} shows the storage space used during query executions obtained from database built-in functions. It is clear that InfluxDB has significantly less memory usage compared to kdb+. In most queries, kdb+ uses 1000 times the memory used for InfluxDB. InfluxDB used the most amount of memory in benchmark O-S, which is still 4 times smaller than kdb+. Previous sections show that kdb+ is overall faster than InfluxDB while it is shown here that InfluxDB is more efficient in memory usage. This result suggests a trade-off between time and space. Moving data and operations to main memory often speed up query execution as accessing main memory is much faster than accessing the disk.

\subsubsubsection{Conclusion}

\begin{itemize}
    \item There exist a trade-off between the time and space used by the query.
\end{itemize}

\begin{table}[H]
    \centering
    \begin{tabular}{ |m{3.2cm}|m{3.2cm}|m{3.2cm}| } 
    \hline
\textbf{Benchmark ID} & \textbf{kdb+ on-disk} & \textbf{InfluxDB}  \\
\hline
 & \multicolumn{2}{c|}{Query Storage (MB)}\\\hline
T-V1     & 109    &  0.016     \\\hline
T-V2    & 83    &  0.0005      \\\hline
T-VWAP   & 48    &  33      \\\hline
O-T   & 2     &  0.003     \\\hline
O-B1   & 0.005     &  0    \\\hline
O-B2   & 0.013     &  0     \\\hline
O-S   & 588     &  145    \\\hline
O-V1   & 184     & 0.694      \\\hline
O-V2   & 184     & 0.153      \\\hline
O-NBBO   & 0.004     &  0.035    \\\hline
C-R   & 109     &  0.014     \\\hline
C-VT   & 54     &  32     \\\hline
C-VO1   & 109     &  0.028    \\\hline
C-VO2   & 109     & 0.023      \\\hline

\end{tabular}
\caption{Performance of Query Storage in kdb+ and InfluxDB}
\label{tab:storage}
\end{table}

\subsection{Benchmark Evaluation}

It can be concluded from the above results and discussion that kdb+ is the most suitable database for financial analysis applications of time series data. kdb+ is able to fast ingest data in bulk, with a good compression ratio for storage efficiency. kdb+ has stable low query latency for all benchmarks including read queries, computational intensive queries, and complex queries. However, kdb+ does occupy significantly larger memory space than InfluxDB when querying.

InfluxDB is the best with storage efficiency, but the data ingestion throughput is significantly slower than kdb+ and TimescaleDB. InfluxDB has the second-best performance in most queries. However, InfluxDB's performance fluctuates for certain operations. It is observed that InfluxDB could have better query latency than kdb+ when the calculation is time window aggregation functions. Additionally, it is suggested that pivoting operations on multiple field columns may significantly slow down the execution.

TimescaleDB is good at data ingestion with a relatively high compression ratio. It has shown to have the largest query latency in most benchmarks. ClickHouse has the worst performance in terms of data ingestion and storage efficiency which could be due to the fact that ClickHouse is not optimised for time series data. It has shown a stable performance of the third best for querying throughout all the benchmarks.

The benchmarking suite designed in this project is shown to be able to evaluate and compare the performance of database systems. Different aspects of the databases including data ingestion, storage efficiency, read query, computationally intensive queries, and complex queries are assessed. The benchmarks use calculations commonly applied to trades and order book data and do not depend on database-specific models or algorithms. This means that this benchmarking suite is easily extensible to other database systems by importing data and executing queries written according to the specification. Additionally, the benchmarking results are not biased by database-specific algorithms.

Additionally, the results from this project are comparable with the results of related works. The conclusion that kdb+ is the most suitable for financial analysis use cases is in line with the results from \parencite{kdb_comparision} and \parencite{kdb_benchmark}. The results obtained on InfluxDB and TimescaleDB in terms of data ingestion, storage efficiency, and query execution are also consistent with \parencite{9458659}.

%%%%%%%%%%%%%%%%%%%%%%%%%%%%%%%%%
\section{Conclusion}
Time series data are measurements of attributes that are recorded in a temporal order. Time series data differs from traditional relational data as it is often large in size, more frequently generated and old data rarely is updated. Many database systems have been developed to optimise for time series data with these characteristics. Time series data has become widely used in the financial industry for use cases such as trading decision-making and market analysis.

Benchmarks can assess database performance and compare the performance of different databases. This project designs a benchmarking suite for financial data analysis use cases extensible to various databases. The benchmarks are implemented and tested on kdb+, InfluxDB, TimescaleDB, and ClickHouse. The collected results are analysed in terms of different aspects of database performance. Additionally, the proposed benchmarking suite is evaluated for its ability to effectively assess database performance.

First, a scenario where time series trades and order book data are stored and queried is defined and relevant use cases are identified. Studies and research related to the use cases are reviewed. The benchmarking suite is designed based on commonly used calculations and metrics summarised from the literature. It covers workload with different data usage, I/O intensity, and computational intensity.

A total of 25 GB of data is collected using the real-time feed from cryptocurrency exchanges over a period of a month. Random data is generated for the bulk writing benchmark. The benchmarking suite is tested on kdb+, InfluxDB, TimescaleDB, and ClickHouse by importing the dataset and implementing the benchmarks in their native query language.

The results of the experiment are analysed, with the conclusion that kdb+ has the overall best performance among the 4 databases. InfluxDB, TimescaleDB, and ClickHouse are also evaluated in various aspects including data ingestion and storage, read queries, computational intensive queries, and complex queries. 

The benchmarking suite is able to show performance differences in different databases and these conclusions are in line with previous works. To our best knowledge, this is the first time series database benchmark designed for a financial analysis use case. Benchmarks in STAC-M3 target financial databases as well but are not publicly accessible.

\subsection{Limitation}

Although the aims and objectives of the project are met, this project exist some limitations. The limitation in the benchmark design lies in the use case. The current design is relatively general and can be applied to a wide array of use cases. For the benchmarking suite to assess database performances more effectively, different benchmarks should be developed targeting uses cases with different data input size and computational complexity. For example, benchmarks for a use case that perform complex pricing models using historical data should vary from benchmarks targeting for fast real-time analysis use case. Additionally, query latency is the only performance metric collected during query execution. Including more metrics could help evaluate database systems more comprehensively.

Limited by the time and memory space of the machine, the data is sampled from 1 month of real-time data. However, many historical data analysis of stock markets include data of a number of assets spanning many months to years. Designing and testing the benchmarks on larger data would allow simulating the calculations more realistically. Another limitation of the hardware is that the benchmarks are only tested in a single thread and with only one client connecting to the server. This means that the benchmarking suite is not able to evaluate the database performance in a scalable setting.

The experimental results could be limited by the database software. This project implemented the benchmarks on selected 4 databases. Additionally, the database software used in this project is all open-source community versions of the software. However, database vendors also provide paid enterprise solutions that include more functionalities. These functionalities may impact database performance and produce different conclusions.

%design limitation: could be more specific to use case (scalable, non scalable, intraday), multi-threading, different data, could use more complex analysis, could get insights from industry

%software limitation: open source vs professional version, some other database not testable

%hardware limitation: PC vs server, cluster, 

%data limitation: only collect sample, not whole dataset, ingestion in multiple formats, dataset schema

%%%%%%%%%%%%%%%%%%%%%%%%%%%%%%%%%%%%
\section{Future Work}
\subsection{Design}

To make the design more specific to use cases, more benchmarks can be developed targeting different use cases. For example, testing a larger-scale implementation could include multiple threads, and many clients connecting to the server at the same time. Scalable systems should be evaluated with clusters of database servers and a large number of client machines.

The sample calculation and analysis used to construct the benchmarks can also be improved. More complex analysis methods can be drawn from research works. Traders and developers in the financial industry could be involved in the design process. They can provide insights on algorithms and data models frequently used in their work and their requirements for the suitable database system under this use case.

\subsection{Experiment}

The experiment can be extended to other databases commonly used for similar workloads. The hardware environment used for the experiment could extend to industry-level servers and clusters. Moreover, server technologies including disk, RAM, and CPU could affect database performance as well. Experimenting on more databases and on more servers provides a comprehensive view of database performance for developers. Developers can choose the suitable database to use considering their scale of implementation, performance preferences and trade-offs, and related costs.

\printbibliography

\end{document}